\documentclass[journal,comsoc]{IEEEtran}

 \usepackage[caption=false,font=footnotesize]{subfig}
\usepackage{graphicx}
\usepackage{amssymb}
\usepackage{amsmath}
\usepackage{amsfonts}
\usepackage{cite}
\usepackage{url}
\usepackage{enumerate}
\usepackage{epstopdf}
\usepackage{dsfont}
\usepackage{suffix}
\usepackage{steinmetz}
\usepackage{amsthm}
\usepackage{algpseudocode}
\usepackage{algorithm}
\usepackage{balance}
\usepackage{color}
\usepackage[mathscr]{euscript}

\newtheorem{corollary}{Corollary}

\newtheorem{proposition}{Proposition}

\DeclareSymbolFont{sfletters}{OML}{cmbrm}{m}{s1}
\DeclareMathSymbol{\sgamma}{\mathord}{sfletters}{"0D}
\DeclareMathSymbol{\sdelta}{\mathord}{sfletters}{"0E}

\DeclareMathSymbol{\seta}{\mathord}{sfletters}{"11}

\begin{document}
%
\title{Multistatic Scatter Radio Sensor Networks \\for Extended Coverage}


\author{Panos N. Alevizos,~\IEEEmembership{Student Member,~IEEE}, 
          Konstantinos Tountas,~\IEEEmembership{Student Member,~IEEE}, \\ 
	 ~and Aggelos Bletsas,~\IEEEmembership{Senior Member,~IEEE}
\thanks{Parts of this work were presented in EURFID 2015 Workshop, Oct. 2015, Rosenheim, Germany. This work was supported by the ERC-04-BLASE project, executed in the context of the
"Education \& Lifelong Learning" Operational Program of the National
Strategic Reference Framework (NSRF), General Secretariat for
Research \& Technology (GSRT), funded through European
Union-European Social Fund and Greek national funds. 
Authors are with Telecom Lab, School of ECE, Technical University of Crete, Chania 73100, Greece. E-mail: {\tt \{palevizos, ktountas\}$@$isc.tuc.gr,  aggelos$@$telecom.tuc.gr}}}


%


\maketitle

\begin{abstract} 
Scatter radio, i.e., communication by means of reflection, has been recently 
proposed as a viable ultra-low power solution for wireless sensor networks (WSNs).
This work offers a detailed comparison between monostatic and multistatic scatter 
radio architectures. In monostatic architecture, the reader consists of both the illuminating
transmitter and the receiver of signals scattered back from the sensors. The multistatic architecture
includes several ultra-low cost illuminating carrier emitters and a single reader.  Maximum-likelihood 
coherent and noncoherent  bit error rate (BER), diversity order, average information and energy outage
probability comparison is performed, under dyadic Nakagami fading, filling a gap in the literature. 
It is found that: (i) diversity order, BER, and tag location-independent performance bounds of 
multistatic architecture outperform monostatic, (ii) energy outage due to   radio frequency  (RF) harvesting for passive 
tags, is less frequent in multistatic than monostatic architecture, and (iii) multistatic coverage 
is higher than monostatic. Furthermore, a proof-of-concept {digital} multistatic, scatter radio WSN
with a single receiver, four low-cost emitters and multiple ambiently-powered, low-bitrate tags, 
perhaps the first of its kind, 
is experimentally demonstrated (at $13$ dBm transmission power), covering an area of $3500$ m$^2$.
Research findings are applicable in the industries of WSNs,    radio frequency identification (RFID), and emerging Internet-of-Things.
\end{abstract}


\begin{IEEEkeywords} Multiple access interference, Nakagami channels, scattering parameters, radio frequency
wireless sensor networks.
\end{IEEEkeywords}

%
\IEEEpeerreviewmaketitle

\section{Introduction}

Typical wireless sensor networks (WSNs) utilize conventional (Marconi) radios for communication. 
Such radio frequency (RF) front-ends typically require signal conditioning units, such as mixers,
active filters, and amplifiers, increasing complexity, power consumption, and cost.  On the other hand, scatter 
radio, i.e., communication by means of reflection, has been recently proposed as a viable ultra-low power
solution for WSNs\cite{VaBlLe:08}. Scattering can 
be achieved with a single RF transistor at each scatter radio node (called tag henceforth), significantly reducing power and monetary cost per WSN terminal. 
Scatter radio has been exploited in various 
interfaces \cite{PaHsB3:01,PaPaHsB3:03} and radio frequency identification
(RFID), widely used for inventorying, electronic tickets, and people identification
\cite{rfid_handbook:03}, \cite{rfid_epc:08}; it is also expected to play a key role in the evolution of Internet-of-Things (IoT) and 
related applications \cite{AlGuMoAlAy:15, Alevizos_PhD_thesis:17}.

In environmental monitoring and precision agriculture, the tag/sensors must 
cover an extended area and thus, link ranges between the tags and the interrogator
(\emph{reader}) must be maximized. There are two different architectures for scatter 
radio networks: the monostatic and the multi-bistatic (multistatic) architectures, depicted in 
Fig~\ref{fig:monostatic_bistatic_topology}. Typical RFID applications utilize the 
monostatic architecture, along with \emph{passive} tags, i.e., powered by the 
illuminating field of the transmitting interrogator. In the monostatic architecture,
the reader consists of both the illuminating transmitter and the receiver of signals 
reflected back from the tags.  Additionally, passive tags require a rectification circuit to convert the RF signals (captured by 
the tag antenna) to DC voltage, powering the tag. Thus, the achievable range is limited by the 
``forward link" \cite{GrDu:09} and the RF harvesting circuitry. Moreover, commercial tags implement
high bitrate communication for the tag-to-reader link, typically in the order of a few hundreds of kilo bits per second, 
resulting in reduced energy per bit, and thus, limited communication range.

In the bistatic architecture, the illuminating carrier emitter (CE) and the receiver
of the reflected (backscattered) signals are distinct units, located at different positions, 
offering flexible network topologies. In the multistatic architecture, several low-cost CEs are available, two orders of magnitude 
cheaper than the reader.  The latter can be a low-cost, commodity software-defined radio (SDR). Due to the morphology of the multistatic
architecture, each  tag can be close to a CE with high probability, offering two desired implications:
(a) the tag-to-reader coverage is increased  with high probability and (b)
using passive tags that harvest RF energy from the illuminating emitters, 
 the probability of energy outage during the energy harvesting phase can be  decreased.

Work in \cite{VaBlLe:08} studied  a semi-passive, low-bitrate scatter radio tag,
i.e., \emph{energy}-assisted tag (e.g., from battery, low-cost solar panel). 
The work studied also the relevant signal processing and noncoherent detection algorithms with scatter radio minimum-shift 
keying (MSK) and monostatic architecture, i.e., with transmitting and receiving antenna 
at the same, custom reader. Frequency division multiplexing among sensors was possible, 
due to frequency-shift keying (FSK) and orthogonal modulating scatter radio frequencies among the tags.
Subsequent work in \cite{BlSiSa:09} studied frequency reuse and adjacent frequency channel 
interference in cellular architectures, where each cell is served by a different reader 
and modulating scatter radio (i.e., \emph{subcarrier}) frequencies are reused across 
different (and distant) cells. Work in \cite{GrDu:07}, \cite{GrDu:08} studied link 
budgets of the general multi-antenna scatter radio link, showing multi-antenna benefits on bit error rate (BER).
Coverage analysis for   WSNs consisting of passive devices  has been conducted in \cite{BeOz:09,HanHu:17}.

Bistatic scatter radio---with emitter and receiver located at different locations,
as separate units---was analyzed and experimentally demonstrated in 
\cite{KiBlSi:12,KiBlSi:13_2, KiBlSa:14}, focusing on noncoherent FSK or
on-off keying and highlighting the idiosyncrasies of scatter radio compared 
to classic, Marconi radio, as well as the additional difficulties imposed by the 
bistatic architecture (e.g., carrier frequency offset between emitter and receiver).
Work in \cite{AlFaTouKaAgBl:14}, \cite{AlBl:15} studied and experimentally demonstrated 
short packet communication with error correction and proposed specific soft-decision
metrics for noncoherent detection in the bistatic scatter radio architecture,
while work in \cite{FaAlBl:15} offered coherent detection and decoding with short
block length channel codes (and short preambles necessary for channel estimation),
after carefully compressing all channel and microwave unknowns of the 
bistatic system to a single complex vector. Coherent detection and SDR-based reception of industrial RFID was shown in \cite{KaMaBl:15}. 
Advanced noncoherent detectors and decoders based on composite hypothesis-testing
are proposed in  \cite{AlBl:15,  AlBlKar:17} performing very close to perfect CSI, emphasizing on short packet communication.
 Scatter radio WSN examples, using, however, analog frequency modulation principles, were given for environmental
 humidity in \cite{KaKiTouKoKoBl:14} and for soil moisture in \cite{DaAsKaBle:14},\cite{DaAsKaBl:16}. Work in \cite{LiuParTalGolWetSmi:13}
 is another example of bistatic scatter radio, where the illuminating signal is the (modulated)
digital television (DTV) signal,
 while work in \cite{IyTaKeGolSm:16} uses bistatic scatter radio principles to convert 
  Bluetooth transmissions to create Wi-Fi and ZigBee-compatible signals. 
Detectors for ambient backscatter systems  relying on bistatic architecture   have been proposed in \cite{WaGaFaTe:16, QiGaWaJiZh:17}


This work proposes multistatic scatter radio WSN architecture for extended
coverage and reliability. A detailed comparison methodology between monostatic and multistatic architectures for scatter radio WSNs is offered, 
examining various performance metrics, such as BER, energy outage probability (in passive tags),
and information outage probability.
The contributions of this work are summarized as follows:
\begin{itemize}
 \item  A multi-user signal model is offered, based  on joint  time- and frequency-division multiplexing (joint TDM-FDM) for both multistatic and monostatic systems, exploiting existing results on scatter radio prior art. The signal model  accounts for (large-scale) path-loss and small-scale fading, based on dyadic Nakagami fading (which can model a wide class of fading channels, including Rayleigh and Rice). 
 \item Upper bounds on BER performance of point-to-point backscatter systems 
 under coherent  maximum-likelihood (ML) detection for dyadic Nakagami fading are derived.
 These bounds coincide with the exact BER performance of noncoherent envelope detection. 
 \item It is found that the bistatic architecture
 has higher diversity order than the monostatic architecture. Additionally, the multistatic analysis covers asymmetric cases, 
 where emitter-to-tag and tag-to-reader links follow different fading statistics. 
  The latter is useful in evaluating recently proposed \emph{ambient} scenarios, 
 where scatter radio tag is close to the receiver but distant from the illuminating ambient emitter.   
\item Information outage probability and tight
Jensen-based upper bounds are provided for Rayleigh fading.
In addition, for  WSNs consisting of passive tags, closed-form expressions for energy outage  are 
derived for Nakagami fading.
\item   A framework for tag topology-independent outage performance evaluation is offered 
based on randomization of tags' positions over square grid topologies. Detailed simulation study reveals that the multistatic architecture outperforms
the monostatic for every studied performance metric.  

 \item A    {digital}
multistatic, scatter radio WSN is experimentally demonstrated and contrasted to a monostatic counterpart, 
corroborating the theoretical findings and offering a concrete proof-of-concept. 
\end{itemize}

The rest of this document is organized as follows: Section~\ref{sec:sys_model} presents the model of
the monostatic and multistatic architectures. Section~\ref{sec:BER} presents the single user 
(i.e., tag) error probability analysis and Section~\ref{sec:outage} offers the outage
probability analysis for multiple users.  
Section~\ref{sec:div_rec_rand_tp}  discusses about diversity
reception and presents how to model the spatial distribution of tags,
while Section~\ref{sec:harvesting} offers
 energy information outage analytical results. 
Numerical and experimental (outdoor) results are presented 
in Sections~\ref{sec:numerical_results} and~\ref{sec:wsn}, respectively. Finally, work is concluded in Section~\ref{sec:conclusion}.  

\emph{Notation:}
Symbols $\mathds{B}$, $\mathds{N}$, $\mathds{R}$, and $\mathds{C}$ denote the set 
of binary, natural, real, and complex numbers, respectively.
$\mathbf{0}_N$ and $\mathbf{I}_N$, denote, respectively, the all-zeros vector
and identity matrix   of size $N$.
The phase of a complex number $z$ is denoted as $\phase{z}$.
The distribution of a proper complex Gaussian $N \times 1$ 
 vector $\mathbf{x}$ with mean $\boldsymbol{\mu}$ and
 covariance matrix $\boldsymbol{\Sigma}$ is denoted by 
$\mathcal{CN}(\boldsymbol{\mu}, \boldsymbol{\Sigma}) \triangleq \frac{1}{\pi^N \mathsf{det}( \boldsymbol{\Sigma})} 
\mathsf{e}^{-(\mathbf{x} - \boldsymbol{\mu})^{\mathsf{H}}   \boldsymbol{\Sigma}^{-1 } (\mathbf{x} - \boldsymbol{\mu})}$.
$\mathcal{U}[a,b)$ denotes the uniform distribution in $[a,b)$. 
 Expectation of function $\mathsf{g}(\cdot)$ of random variable $x$ with
 probability density function (PDF) $\mathsf{f}_{ x}(\cdot)$
 is denoted as ${\mathbb{E}} [ \mathsf{g}({x})] \triangleq
\int_{x}  \mathsf{g}(x)   \mathsf{f}_{ x}( x) \mathsf{d} {x}$.
  The probability of  event $\mathscr{A}$ is denoted as $\mathbb{P}(\mathscr{A})$.

\section{System Model}
\label{sec:sys_model}

\subsection{Network Architecture}
A scatter radio WSN consists of $N$ static sensors (tags) at distinct positions that backscatter their measured data to
a \emph{single} software-defined radio (SDR) reader; the set of tags is denoted as  $\mathcal{N} \triangleq \{ 1, 2, \ldots , N \}$.
Tags/sensors are considered \emph{semi-passive}, i.e., they utilize scatter radio for communication and obtain required power for operation from any type of energy source, either ambient (solar, thermoelectric, chemical) or dedicated (e.g., battery). However, Section~\ref{sec:harvesting} will contrast monostatic and multistatic architectures for \emph{passive} tags as well, which solely harvest energy from the illuminator(s).  

\begin{figure}[t]
\centering
\includegraphics[scale=0.45]{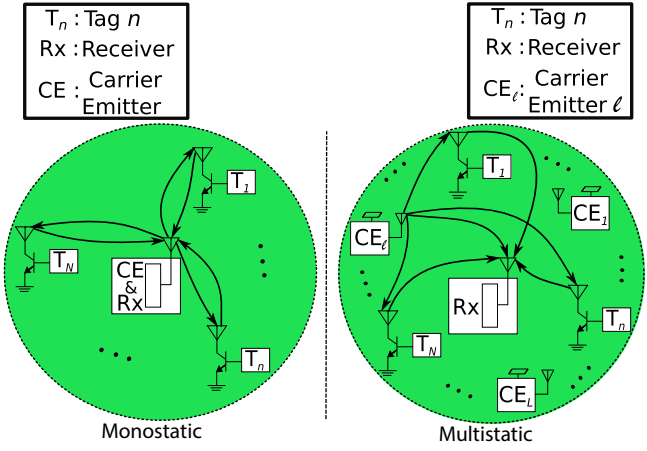}
\caption{Monostatic (Left) and Multistatic (Right) WSN architecture with $N$ tags, $n=1\ldots N$, and $L$ illuminating carrier emitters (CEs), $l=1\ldots L$.}
\label{fig:monostatic_bistatic_topology}
\end{figure}

\begin{figure}[!t]
\centering
\includegraphics[scale=0.55]{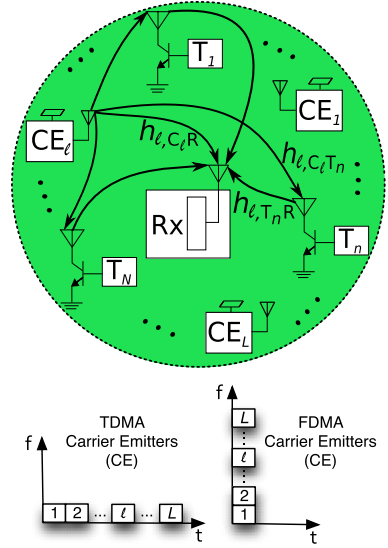}
\caption{Multi-user multistatic  transmission model over the $l$-th time slot.}
\label{fig:bistatic_channel}
\end{figure}

The multistatic network architecture assumes a set of carrier emitters (CE), denoted as $\mathcal{L}\triangleq \{1,2,\ldots, L\}$. The CEs are distinct units from the SDR receiver and transmit a continuous carrier wave (CW) with time-division multiple access (TDMA) or frequency-division multiple access (FDMA) (Fig.~\ref{fig:monostatic_bistatic_topology}-Right and Fig.~\ref{fig:bistatic_channel}). In CEs with TDMA,  the $l$-th CE transmits at the $l$-th \emph{time} slot using a common carrier frequency. In CE FDMA, the $l$-th CE transmits at the $l$-th \emph{frequency} slot, i.e., carrier frequency centered at a frequency band orthogonal to the bands of the other simultaneously (in-time) transmitting CEs.  The wireless channel is assumed quasi-static, changing independently across different (time or frequency) slots. For the $l$-th (time or frequency) slot, there are $2N+1$ unidirectional links  ($N$ CE-to-tag links, $N$ tag-to-SDR reader links and the CE-to-SDR reader link).  The distance between the $l$-th CE and the $n$-th tag  is denoted by
 $ \mathsf{d}_{{\rm C}_{l}{\rm T}_n}$,  the distance between the $n$-th tag and the reader is denoted as
 $ \mathsf{d}_{{\rm T}_n{\rm R}}$ with  $n \in \mathcal{N}$ and the distance between the $l$-th CE and the SDR reader is denoted as $\mathsf{d}_{{\rm C}_{l}{\rm R}}$ 
 (Fig.~\ref{fig:bistatic_channel}). 
 We will show that TDMA and FDMA for CEs offer equivalent signal representations under the assumptions of this work.

The monostatic architecture assumes that the single-antenna reader functions as both the receiver and the CW emitter, 
in full-duplex mode.  Specifically, the reader transmits the illuminating carrier towards the tags, 
which in turn modulate their information on the reflected (and scattered back) signal towards the SDR reader. 
For fair comparison to the multistatic case, transmission across $L$ time slots will be also assumed,  
with wireless channel changing independently between time slots. In this architecture $N$ bidirectional 
links exist, i.e., between the reader and the $N$ tags (Fig.~\ref{fig:monostatic_bistatic_topology}-Left).
The distance between the reader and the $n$-th tag is denoted by $\mathsf{d}_{\rm k}$, where ${\rm k} \in 
\{{\rm T}_n{\rm R}, {\rm R}{\rm T}_n\}$ denotes the unidirectional tag-to-reader, reader-to-tag link, respectively. 

Let $\mathbf{u}_{{\rm T}_n}$, $n \in \mathcal{N}$, $\mathbf{u}_{{\rm C}_l}$, $l \in \mathcal{L}$, and
$\mathbf{u}_{\rm R}$ denote the position of $n$-th tag, $l$-th CE, and SDR reader, respectively. 

\subsection{Channel Model}
\label{subsec:channels}

%
%
The following  path-loss model  is adopted \cite{Goldsmith:05}:
\begin{equation}
\mathsf{L}_{\rm k} =  
   \left(\frac{\lambda}{4 \pi d_0}\right)^2 \left(\frac{d_0}{  \mathsf{d}_{\rm k}}\right)^{\nu_{\rm k}} ,
\label{eq:path_loss}
\end{equation}
where  ${\rm k} \in \left \{  {\rm C}_{l} {\rm R}, {\rm C}_{l}{\rm T}_n ,{\rm T}_n{\rm R}  \right \}$ for multistatic and  
${\rm k} \in \left \{ {\rm T}_n{\rm R} ,  {\rm R}{\rm T}_n   \right \}$ for monostatic architecture, $d_0$ is a reference distance (assumed unit thereinafter), 
 $\lambda$ is the carrier emission wavelength and $\nu_{\rm k}$
is the path-loss exponent for link $\rm k$. 

Tag communication bandwidth and 
channel delay spread are assumed relatively small and thus, frequency non-selective (flat) fading is assumed: complex channel gain for the $l$-th (time or frequency) slot is denoted as $h_{l,\rm k } = a_{l,\rm k}\, \mathsf{e}^{-\mathsf{j} \phi_{l,\rm k}} \label{eq:monostatic_bistatic_channels}$, 
 where $a_{l,\rm k} \in \mathds{R}_+$, $\phi _{l,\rm k}  \in \left[0, 2\pi \right)$, ${\rm k} \in \left \{  {\rm C}_{l} {\rm R}, {\rm C}_{l}{\rm T}_n ,{\rm T}_n{\rm R}  \right \}$ and ${\rm k} \in \left \{
{\rm T}_n{\rm R} ,  {\rm R}{\rm T}_n   \right \}$  for multistatic and monostatic architecture, respectively, with ${\mathbb{E}} \!\left[ |h_{l,\rm k }|^2 \right]  \equiv   {\mathbb{E}} \!\left[ a_{l,\rm k}^2 \right]   =1$. It is emphasized that for CEs in TDMA or FDMA mode, $h_{l_1,\rm k }$ is statistically independent to $h_{l_2,\rm k }$ for any $l_1 \neq l_2$. Moreover, at the monostatic architecture reciprocity implies $h_{l, {\rm T}_n{\rm R}} =  h_{l,  {\rm R}{\rm T}_n}$,
while at the multistatic architecture, $\{h_{l,\rm k }\}$ are independent  (and not necessarily identically distributed) for  different ${\rm k} \in \left \{  {\rm C}_{l} {\rm R}, {\rm C}_{l}{\rm T}_n ,{\rm T}_n{\rm R}  \right \}$.\footnote{Results can be easily extended to the multi-antenna reader case.}

Due to potentially strong line-of-sight (LoS) signals in scatter radio environments, Nakagami small-scale 
fading is assumed (with ${\mathbb{E}} \!\left[ a_{l,\rm k}^2 \right]   =1$) \cite[p.~79]{Goldsmith:05}:
\begin{equation}
 \mathsf{f}_{a_{l,\rm k}}(x) = 2\,(\mathtt{M}_{\rm k})^{\mathtt{M}_{\rm k}} \, \frac{x^{2   \mathtt{M}_{\rm k}- 1}}{\Gamma(\mathtt{M}_{\rm k})} \, \mathsf{e}^{-\mathtt{M}_{\rm k} x^2}, ~~x \geq 0 ,
\label{eq:Nakagami_fading}
\end{equation}
where  $\mathtt{M}_{\rm k} \geq \frac{1}{2}$ is the Nakagami parameter and  function $\Gamma(x)= \int_{0}^{\infty}t^{x-1} \mathsf{e}^{-t} \mathsf{d}t$ 
is the Gamma function. For the special cases of $\mathtt{M}_{\rm k} =1$ and  $\mathtt{M}_{\rm k} =\infty$, 
Rayleigh (i.e., NLoS scenarios) and    zero fading  (i.e., $ a_{l,\rm k} = 1$)  is  obtained, respectively. For $\mathtt{M}_{\rm k}= \frac{(\kappa_{\rm k} + 1)^2}{2\kappa_{\rm k}  + 1}$, the distribution in~\eqref{eq:Nakagami_fading} approximates Rice with 
Rician parameter $\kappa_{\rm k} $,    where commonly
$\kappa_{\rm k} \in  [0, 20]$  \cite{Goldsmith:05}.\footnote{  
For the calculation of bit error rate under noncoherent detection, 
phases $\phi _{l,\rm k}$ are assumed  uniformly distributed  over $[0,2 \pi)$.
The expressions for the rest  metrics  are independent of the distribution of phases $\phi _{l,\rm k}$.}
It is noted that the adopted small-scale fading   model can accommodate either LoS or NLoS scenarios by alternating 
the value of  Nakagami parameter $\mathtt{M}$ for each scatter radio link.

\subsection{Signal Model}

\subsubsection{Multistatic}
\label{subsec:bistatic_WSN}

The $l$-th CE transmits a CW at the $l$-th (time or frequency) slot, with complex baseband
$
{q}^{[\rm b]}_{{l}}(t) = \sqrt{2 P_{{\rm C}_l}} \, \mathsf{e}^{-\mathsf{j}(2 \pi \Delta F_l t + \Delta \phi_l)}
$, where $ P_{{\rm C}_l}$ is the $l$-th CE transmission power, while $\Delta F_l$ and 
$\Delta \phi_l$ model the carrier frequency offset (CFO) and phase offset between the $l$-th CE and the
 SDR reader, respectively, due to the fact that CE and the SDR reader do not share the same oscillator.

Each tag $n \in \mathcal{N}$ is illuminated by the carrier wave ${q}^{[\rm b]}_{{l}}(t)$. 
 The tag's information is binary-modulated on the incident CW 
 by switching the antenna load between two loads, associated with two distinct 
 reflection coefficients: $\Gamma_{n, 0}$ and $\Gamma_{n, 1}$ for bit ``0'' and bit ``1'', respectively. For scatter 
 radio, binary frequency-shift keying (FSK) is assumed in this work, where each tag switches between the two loads using a 
 $50 \%$ duty cycle square waveform of duration $T$ per bit (nominal  bit duration), fundamental frequency
 $F_{n, i}$, and random initial phase $\Phi_{n, i }$, 
for  bit $i  \in \mathds{B}\triangleq \{0, 1\}$; in other words, modulation occurs with different 
switching frequency between tag antenna loads, utilizing switching (also coined as \emph{subcarrier}) frequency $F_{n, 0}$ for bit ``0'' 
and $F_{n, 1}$  for bit ``1'', without any type of signal conditioning, such as filtering, amplification or mixing \cite{VaBlLe:08}. 

\begin{figure}[!t]
\centering
\includegraphics[scale=0.37]{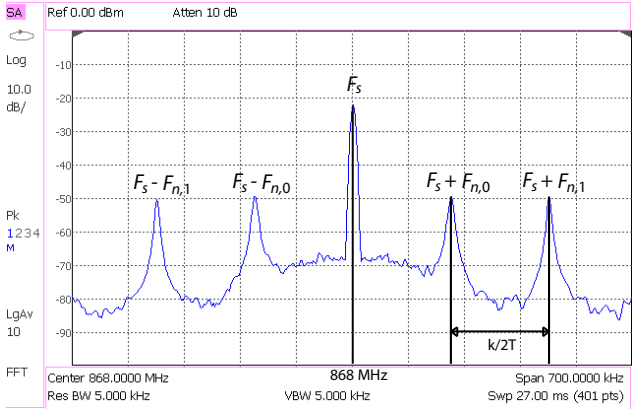}
\caption{Measured backscatter radio FSK spectrum with two loads. Scattered signal appears in $4$ frequencies, $2$ left and $2$ right of the illuminating carrier.}
\label{fig:tag_spectrum}
\end{figure}

Fig.~\ref{fig:tag_spectrum} shows the measured spectrum at a spectrum analyzer when a signal's generator's CW at $F_s=868$ MHz illuminates
tag $n$ that utilizes binary FSK with $|F_{n, 1}-F_{n, 0}|=122.5$ kHz.  In contrast to classic (Marconi-radio) FSK, frequencies
$F_s-F_{n,0}$ and $F_s-F_{n,1}$ also contribute to the total signal (in addition to $F_s+F_{n,0}$ and $F_s+F_{n,1}$)\cite{VaBlLe:08} and thus,
backscatter FSK modulation uses $4$ frequencies, $\pm F  _{n, i_n} , i_n \in \mathds{B}$.
Therefore,  the optimal receiver requires $4$ matched filters and not $2$, as in classic 
FSK demodulation.

Each tag  $n$ can use a \emph{unique} set of subcarrier
 frequencies $F_{n, i_n}, i_n \in \mathds{B} $ to scatter its information. 
At the network level,  the $4N$ frequencies 
$\left\{\pm F  _{n, i_n} \right\} , \forall (i_n,n) \in \mathds{B} \times \mathcal{N} $ must satisfy the  orthogonality criterion, 
which for coherent FSK adheres to the following:%
\footnote{For noncoherent FSK, term $k/2T$ in Eq.~\eqref{eq:criterion_non_coh} is replaced with $k/T$. }%
\begin{equation}
\left| F_{n, i} - F_{j,  m} \right| = \frac{k }{2T} \quad \text{ and} \quad
F_{n, i} \gg \frac{1}{2T},
\label{eq:criterion_non_coh}
\end{equation}
 $\forall (i,n)$, $(m, j) \in   \mathds{B} \times \mathcal{N}  : m \neq i$, and  $k \in \mathds{N}$.
 
This  implies that  the system of $N$ tags scattering simultaneously can be divided into 
 $N$ parallel orthogonal channels, where each can be received separately, without collision. Such single-tag processing techniques have been extensively
 covered in \cite{VaBlLe:08,KiBlSa:14,FaAlBl:15,AlBl:15}. At the $l$-th slot,  the 
CFO-free, DC-blocked received baseband signal at the reader over a bit period ${T}$ from tag $n$, excluding higher harmonics, is given by 
\cite[Eq.~(13)]{FaAlBl:15}, \cite[Eq.~(12)]{AlBl:15}: 
\begin{equation}
\mathbf{r}^{[\rm b]}_{l,n}  = h^{[\rm b]}_{l,n} \, \sqrt{\mathtt{E}^{[\rm b]}_{l,n}} \, \mathbf{x}_{n} + \mathbf{w}_{l,n}, \quad n \in \mathcal{N},
\label{eq:basevand_equivalent_bistatic}
\end{equation}
 where the vector $\mathbf{x}_{n}$ is defined as: 
\begin{equation}
\mathbf{x}_{n} \triangleq  \sqrt{\frac{1}{2}}
\left[  
\mathsf{e}^{+\mathsf{j} \Phi _{n,0} }~
\mathsf{e}^{-\mathsf{j} \Phi _{n,0} }~
\mathsf{e}^{+\mathsf{j} \Phi _{n,1} }~
\mathsf{e}^{-\mathsf{j} \Phi _{n,1} }~
 \right]^{\top}
\odot \mathbf{v}_{ i_n},  
\label{eq:t_i_n}
\end{equation}
$ n \in \mathcal{N},$ where $\mathbf{v}_{i_n} = \left[ (1 - i_{n}) \,~ (1 - i_{n}) \,~ i_{n} \,~ i_{n} \right]^{\top}$ is the four-dimensional 
transmitted symbol of the $n$-th tag corresponding to transmitted bit $i_{n} \in \mathds{B}$. $\Phi _{n,0}$   ($\Phi_{n,1}$) is implementation-specific phase mismatch between tag $n$
and reader for bit $0$   (bit $1$), assumed  
constant for the  $L$ slots. Symbol $\odot$ denotes the  component-wise (Hadamard) product. For  $F_{n,i_n}+ \frac{20}{T} \ll W_{\rm SDR}$, where $W_{\rm SDR}$ is the SDR receiver baseband bandwidth,  
$\mathbf{w}_{l,n} \sim\mathcal{CN}\left( \mathbf{0}_4, 
{N}_0  \, \mathbf{I}_4 \right)$  \cite[Theorem~1]{FaAlBl:15},  with $N_0 =\mathrm{k_b} \mathrm{T_{\theta}}/2$, where $\mathrm{k_b}$ and $ \mathrm{T_{\theta}}$
are the Boltzmann constant and receiver temperature, respectively.  $h^{[\rm b]}_{l,n}$ in Eq.~\eqref{eq:basevand_equivalent_bistatic} is given by: 
\begin{align}
h^{[\rm b]}_{l,n}&\triangleq {a}^{[\rm b]}_{l,n}  \,  \mathsf{e}^{-\mathsf{j} \phi^{[\rm b]}_{l,n} }, \\
 {a}^{[\rm b]}_{l,n}&= a_{l,{\rm C}_{l}{\rm T}_n}   \, a_{ l,{\rm T}_n {\rm R} },  \label{eq:magnitude_bistatic} \\
 \phi^{[\rm b]}_{l,n}&=   \phi _{l,{\rm C}_{l}{\rm T}_n} + \phi _{l, {\rm T}_n {\rm R} } + \Delta \phi_l  
+ \phase{ \Gamma _{n,0} - \Gamma _{n,1} }.
\end{align}
Symbol  $ \mathtt{E}^{[\rm b]}_{l,n}$ denotes the   average energy per bit for the $n$-th tag over the $l$-th slot 
given by \cite[Eq.~(9)]{AlBl:15}:
\begin{equation}
\mathtt{E}^{[\rm b]}_{l,n} 
=     \frac{ {\mathbb{E}} \!\left[    \left( {a}^{[\rm b]}_{l,n} \,\mu^{[\rm b]}_{l,n} 
\right)^2  T \right]  }{2}    = \frac{ \left(\mu^{[\rm b]}_{l,n} 
\right)^2  T  }{2},  
\label{eq:avg_energy_bistatic}
\end{equation}
taking into account that RVs $ a_{l,{\rm C}_{l}{\rm T}_n} $ and $a_{ l,{\rm T}_n {\rm R} }$
are independent with unit average squared value. The following is also employed:
\begin{equation}
\mu^{[\rm b]}_{l,n} =    \sqrt{ 2\, P_{{\rm C}_l} \,\mathsf{L}_{{\rm C}_{l}{\rm T}_n}\, \mathsf{L}_{ {\rm T}_n {\rm R} }}  \,
\left| \Gamma _{n,0} - \Gamma _{n,1} \right| \, \frac{2}{\pi} \, \mathtt{s}_n,
\label{eq:mu_bistatic}
\end{equation}
 with $\mathtt{s}_{n}$ modeling the $n$-th tag's  (real) scattering efficiency, assumed constant. The average received SNR of tag $n$ at the $l$-th  slot  
for multistatic system associated with the system in Eq.~\eqref{eq:basevand_equivalent_bistatic} is given by 
$\mathtt{SNR}^{[\rm b]}_{l,n}  = \mathtt{E}_{l,n}^{[\rm b]}/N_0.
\label{eq:average_received_SNR_bistatic}
$

As already mentioned, for CEs in TDMA or FDMA mode, $h_{l_1,\rm k }$ is statistically independent to $h_{l_2,\rm k }$ for any $l_1 \neq l_2$. 
Under this assumption, tools developed in this work can in principle accommodate both CE modes. For simpler presentation and concise comparison 
with the monostatic architecture, we will assume hereinafter CEs in TDMA mode.

\subsubsection{Monostatic}
\label{subsec:monostatic_WSN}

Due to the fact that the receiver and the emitter share the same 
oscillator, CFO $\Delta F_l$ and phase offset $\Delta \phi_l$ are zero in the monostatic case. Using  Eq.~\eqref{eq:basevand_equivalent_bistatic} and the reciprocity of the channel between the reader and each tag,
 the DC-blocked, demodulated received signal for tag $n$ in the monostatic architecture at time slot $l$ is given by:  
\begin{equation}
\mathbf{r}^{[\rm m]}_{l,n} = h^{[\rm m]}_{l,n}\, \sqrt{\frac{\mathtt{M}_n}{\mathtt{M}_n + 1} \mathtt{E}^{[\rm m]}_{n}} \, \mathbf{x}_{  n} + \mathbf{w}_{l,n}, \quad n \in \mathcal{N},
\label{eq:baseband_equivalent_monostatic}
\end{equation}
under the same assumptions and definitions as in Eqs.~\eqref{eq:basevand_equivalent_bistatic},~\eqref{eq:t_i_n}. In accordance with the multistatic case, $h^{[\rm m]}_{l,n}$ is given by: 
\begin{align}
 h^{[\rm m]}_{l,n} & \triangleq {a}^{[\rm m]}_{l,n} \, \mathsf{e}^{-\mathsf{j} \phi^{\rm [m]}_{l,n}}, \\
{a}^{[\rm m]}_{l,n}&= \left(a_{l, {\rm T}_n {\rm R} } \right)^2, ~\phi^{[\rm m]}_{l,n}=   2 \phi _{ l,{\rm T}_n {\rm R} }  
+ \phase{ \Gamma_{n,0} - \Gamma_{n,1} }.
\end{align}
The average received energy per bit  for monostatic system  for the $n$-th tag over the $l$-th time slot, $\mathtt{E}^{[\rm m]}_{n}$, is expressed as:
\begin{equation}
\mathtt{E}^{[\rm m]}_{n} =    \frac{  \mathbb{E}  \left[ \left( {a}^{[\rm m]}_{l,n} \,\mu^{[\rm m]}_{l,n} 
\right)^2  T \right] }{2}    =  \frac{ 1 + \mathtt{M}_{n} }{2   \, \mathtt{M}_n }\left(\mu^{[\rm m]}_{ n}
\right)^2  T,
\label{eq:avg_energy_monostatic}
\end{equation}
since ${\mathbb{E}} \!\left[ 
\left( a^{[\rm m]}_{l,n}\right)^2\right] = {\mathbb{E}} \!\left[ 
\left( a_{ l,{\rm T}_n {\rm R} }\right)^4\right] = (\mathtt{M}_n +1)/\mathtt{M}_n$ and $\mu^{[\rm m]}_{ n}$ is simplified from Eq.~\eqref{eq:mu_bistatic} to:
\begin{equation}
\mu^{[\rm m]}_{ n} =   \sqrt{2 P_{\rm R}}  \,  \mathsf{L}_{ {\rm T}_n {\rm R} }  
\left| \Gamma_{n,0} - \Gamma_{n,1} \right| \, \frac{2}{\pi} \, \mathtt{s}_n. 
\label{eq:mu_monostatic}
\end{equation}
The average received SNR of tag $n$ at the $l$-th time slot  
for monostatic system is given by\footnote{From Eq.~\eqref{eq:baseband_equivalent_monostatic}, notice that ${\mathbb{E}} \!\left[ |h^{[\rm m]}_{l,n}|^2 \frac{\mathtt{M}_n}{\mathtt{M}_n + 1}  \mathtt{E}^{[\rm m]}_{n} ||\mathbf{x}_n||^2 \right] = \mathtt{E}^{[\rm m]}_{n}$.}
$
\mathtt{SNR}^{[\rm m]}_{n}  = \mathtt{E}^{\rm [m]}_{ n}/N_0.
\label{eq:average_received_SNR_monostatic}
$

It is emphasized that the quantities $\mathtt{E}^{[\rm m]}_{n}$
and $\mathtt{SNR}^{[\rm m]}_{n} $ in the monostatic case above, do not depend on time index $l$,
because they are functions of path-loss  $\mathsf{L}_{ {\rm T}_n {\rm R} } $; the latter  remains   unaffected during the $L$ slots.
In contrast,  $\mathtt{E}^{[\rm b]}_{l,n}$
and $\mathtt{SNR}^{[\rm b]}_{l,n} $ in the multistatic case,
  are both functions of  path-loss $ \mathsf{L}_{{\rm C}_{l}{\rm T}_n} $, which depends on the slot index $l$, since different CE corresponds to  each slot.
\label{remark:SNR_mon_vs_bis}

In addition, it is noted that due to the   nature of FSK  modulation,
 the discrete  baseband  signal expressions in  Eqs.~\eqref{eq:basevand_equivalent_bistatic}
and~\eqref{eq:baseband_equivalent_monostatic} depend solely on  links  $ {\rm C}_{l}{\rm T}_n$
and ${\rm T}_n{\rm R} $   due to DC-blocking operation, and thus,  the analysis is continued with  link ${\rm T}_n{\rm R} $
for monostatic and links  $ {\rm C}_{l}{\rm T}_n$
and ${\rm T}_n{\rm R} $ for bistatic (one CE) or multistatic (multiple CEs) architecture, $\forall (l,n) \in \mathcal{L} \times \mathcal{N}$.

%
%

\section{BER Analysis with ML Detection}
\label{sec:BER}

\subsection{Distribution of Fading Amplitudes}

It can be observed that RVs  $(a_{l,  { {\rm C}_{l}{\rm T}_n}})^2$ and $(a_{l,  { {\rm T}_n{\rm R}  }})^2$   
are independent following Gamma distribution  with parameters 
$\left(\mathtt{M}_{{\rm C}_{l}{\rm T}_n}, \frac{1 }{\mathtt{M}_{{\rm C}_{l}{\rm T}_n}} \right)$
and $\left(\mathtt{M}_{{\rm T}_n{\rm R}}, \frac{1 }{\mathtt{M}_{{\rm T}_n{\rm R}} }\right)$,
respectively \cite[p.~242]{pappoulis:02}, i.e.,  the  probability density function (PDF)  
of RV $(a_{l,\rm k})^2$ is given by:
\begin{equation}
 \mathsf{f}_{a_{l,{\rm k} }^2}(x)   =  (\mathtt{M}_{\rm k})^{\mathtt{M}_{\rm k}} \,
 \frac{x^{ \mathtt{M}_{\rm k}- 1}}{\Gamma(\mathtt{M}_{\rm k})} \, \mathsf{e}^{-\mathtt{M}_{\rm k} x }, ~~x \geq 0, 
\label{eq:gamma_distr}
\end{equation}
 ${\rm k} \in \left \{   {\rm C}_{l}{\rm T}_n ,{\rm T}_n{\rm R}  \right \}$. The above distribution reflects
the power distribution of each small-scale fading scatter radio link for both monostatic and multistatic architectures
and will be used as a building block to derive closed-form expressions for the metrics of interest.

For simplified notation, the following abbreviations are used: $\mathtt{M} _{ {\rm C}_l{\rm T}_n} = \mathtt{M}_{ln}$ and 
$\mathtt{M}_{ {\rm T}_n {\rm R}} = \mathtt{M}_n$.

\subsection{Coherent}

\subsubsection{Monostatic}

According to Eq.~\eqref{eq:baseband_equivalent_monostatic}, the conditional bit error rate (BER) 
for tag $n$ over time slot $l$,  given channel realization $h^{\rm[ m]}_{l,n}$ (with $\left| h^{[\rm m]}_{l,n}\right| =
a^{\rm [m] }_{l,n}$) and phase parameters $\Phi_{n,0}, \Phi_{n,1}$  
under ML coherent detection  depends solely  on  amplitude $a^{\rm [m] }_{l,n}$ \cite{FaAlBl:15}, and is given by  \cite[p.~508]{TsVi:05}:
\begin{align}
\mathbb{P} \! \left(e_{l,n}^{\rm [m]} \mid a^{[\rm m]}_{l,n} \right) &= \mathsf{Q}\!\left( \frac{a^{\rm [m] }_{l,n}
\sqrt{\frac{\mathtt{M}_n}{\mathtt{M}_n + 1}\mathtt{E}^{\rm [m]}_{n}} \,\left \| \mathbf{x}_0- \mathbf{x}_1\right  \|_2}
{\sqrt{2 \,N_0}} \right) \nonumber \\ &=
\mathsf{Q}\!\left( a^{[\rm m]}_{l,n}  \sqrt{ \frac{\mathtt{M}_n  \mathtt{SNR}^{[\rm m]}_{n}}{\mathtt{M}_n + 1}} \right),
\label{eq:coherent_basic_monostatic}
\end{align} 
 where function $\mathsf{Q}(x) = \frac{1}{2\pi} \int_x^\infty \mathsf{e}^{-\frac{t^2}{2}}\mathsf{dt}$ and 
 $\left \| \mathbf{x}_0- \mathbf{x}_1\right \|_2  =\sqrt{ 2} $ were utilized. 
 
Using the Chernoff bound for $\mathsf{Q}(\cdot)$ function,  Eq.~\eqref{eq:coherent_basic_monostatic} can be upper bounded
 as 
$
 \mathsf{Q}\!\left( a^{[\rm m]}_{l,n}  \sqrt{ \frac{\mathtt{M}_n  \mathtt{SNR}^{[\rm m]}_{n}}{\mathtt{M}_n + 1}} \right)
\! \leq\! \frac{1}{2}\,\mathsf{e}^{  - \frac{\left(  a^{[\rm m]}_{l,n}  \right)^2   \, \mathtt{M}_n  \, \mathtt{SNR}^{[\rm m]}_{n}}{2\,(\mathtt{M}_n + 1) }  } $.
Since  RV $ {a}^{[\rm m]}_{l,n} = \left( a_{ l,{\rm T}_n {\rm R} }\right)^2$ follows Gamma 
distribution with  parameter $\left(\mathtt{M}_n, \frac{1}{\mathtt{M}_n} \right)$, the unconditional BER can be bounded as follows:
 \begin{align}
\! \mathbb{P} \!\left(e^{\rm [m]}_{l,n} \right) \!&= \!\underset{a^{[\rm m]}_{l,n}}{\mathbb{E}} 
\left[ \mathbb{P} \! \left( e^{\rm [m]}_{l,n} \mid a^{[\rm m]}_{l,n}\right) \right]\nonumber \\
&\overset{(a)}{\leq}
\!\int_0^{\infty} \! \frac{1}{2}\,\mathsf{e}^{  - \frac{x^2    \, \mathtt{M}_n  \, \mathtt{SNR}^{[\rm m]}_{n}}{2\,(\mathtt{M}_n + 1) }  }
 (\mathtt{M}_{n})^{\mathtt{M}_{n}} \,
 \frac{x^{ \mathtt{M}_{n}- 1}}{\Gamma(\mathtt{M}_{n})} \, \mathsf{e}^{-\mathtt{M}_{n} x } \mathsf{d}x \nonumber \\
&\overset{(b)}{=} \frac{1 }{2}    \left( \frac{\mathtt{M}_{n}+\mathtt{M}_{n}^2}{2 \, \mathtt{SNR}^{[\rm m]}_{n}} \right)^{\!\frac{\mathtt{M}_{n}}{2}} 
 \mathsf{U}\!\left(\frac{\mathtt{M}_{n}}{2}, \frac{1}{2}, \frac{\mathtt{M}_{n}+\mathtt{M}_{n}^2}{2\, \mathtt{SNR}^{[\rm m]}_{n}} \right),
\label{eq_prob_error_monostatic_upper_bound}
\end{align} 
where $ \mathsf{U}( \cdot, \cdot, \cdot) $ is the 
confluent hypergeometric function \cite[Eq.~(13.4.4)]{Olver:10}. In step $(a)$ above, the Chernoff bound for 
$\mathsf{Q}$ function was exploited and in step $(b)$, \cite[Eq.~(3.462.1) and (9.240)]{gradshteyn:07} and then \cite[Eq.~(13.14.3)]{Olver:10} were utilized to simplify the final formula. 

Is is emphasized that BER of Eq.~\eqref{eq_prob_error_monostatic_upper_bound} depends on  $\mathtt{SNR}^{[\rm m]}_{n}$ of 
tag $n$, which in turn, depends on Eq.~\eqref{eq:avg_energy_monostatic}; the latter is a function of tag's $n$ location through
Eq.~\eqref{eq:mu_monostatic}. Thus, the above BER expression depends on the topology of the tags. The following proposition offers an important, topology-independent metric:

\begin{proposition} \normalfont
Under Rayleigh fading, i.e.,  $\mathtt{M}_n= 1$,   monostatic architecture offers diversity order equal to $1/2$ for any $(l,n) \in \mathcal{L} \times \mathcal{N}$.
\begin{IEEEproof}
The proof is given in Appendix~\ref{Appendix:app1}.   
\end{IEEEproof}
\label{prop:monostatic_div_order}
\end{proposition}

The above result indicates that for any slot under Rayleigh fading, monostatic BER decays inversely proportional with
square root of  SNR at the high SNR regime. It is shown below that the decay is faster in the multistatic case. 

\subsubsection{Multistatic}

Exploiting Eq.~\eqref{eq:magnitude_bistatic} along with the formula in \cite[p.~302, Eq.~6.148]{pappoulis:02},
the PDF of the product $ {g}^{[\rm b]}_{l,n}\triangleq \left({a}^{[\rm b]}_{l,n}\right)^2 = (a_{l,  { {\rm C}_{l}{\rm T}_n}})^2  ~ (a_{l,  { {\rm T}_n{\rm R}  }})^2$
can be expressed in closed form as follows ($x\geq 0$):
\begin{align}
\mathsf{f}_{ {g}^{[\rm b]}_{l,n}}(x)  &= \int_{0}^{\infty} \frac{1}{y} \,
\mathsf{f}_{ a_{l,{\rm C}_{l}{\rm T}_n}^2}\!\!(y)\, \mathsf{f}_{a_{ l,{\rm T}_n {\rm R} }^2}\!\!\left(\frac{x}{y}\right) \mathsf{d}y  \nonumber \\
&=
  \frac{ 2 ( x \, \mathtt{M}_{ ln}  \, \mathtt{M}_{n} )^{\frac{\mathtt{M}_{ln } + \mathtt{M}_n}{2}}   \,
\mathsf{K}_{ \mathtt{M}_{n} - \mathtt{M}_{ln }  }\! \left( 2 \sqrt{\mathtt{M}_{ ln}  \, \mathtt{M}_{n} \, x} \right)
  } { x\, \Gamma(\mathtt{M}_{ln})\, \Gamma(\mathtt{M}_{n})}, 
\label{eq:bistatic_power_distr}
\end{align}
where \cite[Eq.~(3.471.9)]{gradshteyn:07} was used to obtain the simplified form in Eq.~\eqref{eq:bistatic_power_distr}.
 $\mathsf{K}_{\nu}(\cdot)$ is the $\nu$-th order modified Bessel function of the second kind,
satisfying  $\mathsf{K}_{\nu}(\cdot) =  \mathsf{K}_{-\nu}(\cdot)$ \cite[Eq.~(10.27.3)]{Olver:10}.
The above distribution is the power distribution of Nakagami dyadic backscatter channel.  Similar expression with Eq.~\eqref{eq:bistatic_power_distr}  can be found in \cite{ChaKaMich:09}, while a
derivation of~ Eq.~\eqref{eq:bistatic_power_distr}  in the special case of  $\mathtt{M}_{ ln} = \mathtt{M}_{ n}$ is given in \cite{BiSaTsKa:07}.

In the multistatic case, according to Eq.~\eqref{eq:basevand_equivalent_bistatic},
the conditional BER  for tag $n$ over the $l$-th time slot is given by: 
\begin{equation}
\mathbb{P} \! \left(e^{\rm [b]}_{l,n}  \mid a^{[\rm b]}_{l,n}\right)
 =   \mathsf{Q}\!\left(  a^{[\rm b]}_{l,n} \,   \sqrt{\mathtt{SNR}^{[\rm b]}_{l,n}} \right) ,
 \label{eq:cond_prob_error_bistatic}
\end{equation}
which with the use of Chernoff bound is upper bounded as $ \mathsf{Q}\!\left(  a^{[\rm b]}_{l,n} \, 
\sqrt{\mathtt{SNR}^{[\rm b]}_{l,n}} \right) \leq\! \frac{1}{2}\,\mathsf{e}^{  - \frac{\left( a^{[\rm b]}_{l,n}  \right)^2   \, \mathtt{SNR}^{[\rm b]}_{l,n}}{2} }.$ 

Hence, the BER over the $l$-th slot for the $n$-th tag  is upper bounded as follows:
\begin{align}
& \mathbb{P} \!\left(e^{\rm [b]}_{l,n} \right)  = 
\underset{a^{[\rm b]}_{l,n}}{\mathbb{E}}\!\! \left[ \mathbb{P} \! \left(e^{\rm [b]}_{l,n}  \big| a^{[\rm b]}_{l,n}  \right) \right]  
\nonumber \\
& \overset{}{\leq} \!  \int_{0}^{\infty} \!  \mathsf{e}^{\left( -
 x   \mathtt{SNR}^{[\rm b]}_{l,n} / 2 \right)}
 \frac{   (  x \, \mathtt{M}_{ ln}  \, \mathtt{M}_{n} )^{\frac{\mathtt{M}_{ln } + \mathtt{M}_n}{2}}   \,
\mathsf{K}_{ \mathtt{M}_{n} - \mathtt{M}_{ln }  }\! \left( 2 \sqrt{\mathtt{M}_{ ln}  \, \mathtt{M}_{n} \, x} \right)
  } {   x\, \Gamma(\mathtt{M}_{ln})\, \Gamma(\mathtt{M}_{n})}  \mathsf{d} x \nonumber
\\
&\overset{(a)}{=}
\frac{1}{2} \left(\frac{2\, \mathtt{M}_{ ln}  \mathtt{M}_{n}}{  \mathtt{SNR}^{[ \rm b]}_{l,n}}\right)^{\!\! \mathtt{M}_{n}}
 \mathsf{U}\!\left(\mathtt{M}_{n}, 1 + \mathtt{M}_{n} - \mathtt{M}_{ln}, \frac{2\, \mathtt{M}_{ ln}  \mathtt{M}_{n}}{  \mathtt{SNR}^{[\rm b]}_{l,n}} \right),
\label{eq:proof_probability_of_error_bistatic}  
\end{align}
where $ \mathsf{U}( \cdot, \cdot, \cdot) $ is the 
confluent hypergeometric function \cite[Eq.~(13.4.4)]{Olver:10}. In Step $(a)$ above, change of variables $x = y^2$ is applied and then \cite[Eq.~(6.631.3)]{gradshteyn:07} 
and \cite[Eq.~(13.14.3)]{Olver:10} are exploited to simplify the final expression.

BER of Eq.~\eqref{eq:proof_probability_of_error_bistatic} also depends on the network topology, through the definition of  $\mathtt{SNR}^{[ \rm b]}_{l,n}$
 and energy per bit per slot in Eqs.~\eqref{eq:avg_energy_bistatic} and~\eqref{eq:mu_bistatic}. The following proposition offers a topology-independent metric:
\begin{proposition} \normalfont
Under Rayleigh fading, i.e., $ \mathtt{M}_{ ln} = \mathtt{M}_{n}=1$, multistatic architecture offers diversity order \emph{at least} $1$
 for any $(l,n) \in \mathcal{L} \times \mathcal{N}$.
\begin{IEEEproof}
The proof is given in Appendix~\ref{Appendix:app2}.
\end{IEEEproof}
\label{prop:bistatic_div_order}
\end{proposition}

Thus, it is concluded that under Rayleigh fading, the multistatic BER drops faster compared to the monostatic, 
at the high SNR regime, even for a single, fixed slot.

\subsection{Noncoherent}

To better highlight the importance of the expressions derived in
Eqs.~\eqref{eq_prob_error_monostatic_upper_bound} and~\eqref{eq:proof_probability_of_error_bistatic}, consider the $n$-th tag operating in noncoherent
reception mode over the $l$-th slot.
For fixed dyadic backscatter channel 
  amplitudes  $a^{[\rm m]}_{l,n}$ and $a^{[\rm b]}_{l,n}$, as well as phase offsets $\Phi_{n,0}$ and $\Phi_{n,1}$,
  and unknown  dyadic backscatter channel angles  $\phi^{[\rm m]}_{l,n}$ and $\phi^{[\rm b]}_{l,n}$,
  it is not difficult to show that
the ML reception  rule for orthogonal signaling is based on envelope detection \cite[Eq. 4.5-32]{PrSal:07}
of the $4 \times 1$ complex received vector $\mathbf{r}_{l,n}$ and takes the following form: 
\begin{align}
& \hspace{-6pt}\left|   \mathbf{r}^{\rm [x]}_{l,n}[1] +  
  \mathsf{e}^{ 2\mathsf{j} \Phi_{n,0} } \mathbf{r}^{\rm [x]}_{l,n}[2]  \right| \!
   \underset{i_n=1}{\overset{i_n=0}{\gtrless}}
\!  \left|  \mathbf{r}^{\rm [x]}_{l,n}[3] +  
  \mathsf{e}^{ 2 \mathsf{j} \Phi_{n,1} } \mathbf{r}^{\rm [x]}_{l,n}[4]  \right|,
  \label{eq:non_coh_detec_rule}
\end{align}
where $\rm x = m$, $\rm x = b$, for monostatic and multistatic system, respectively.
Note that the above rule  requires the   received signal vector 
$\mathbf{r}^{\rm [x]}_{l,n}$ and the tag $n$ implementation-specific phases $\Phi_{n,0}$ and $\Phi_{n,1}$, while it is different than the square-law detector:  
$
| \mathbf{r}_{l,n}[1] |^2  +   
  | \mathbf{r}_{l,n}[2]  |^2 \!
   \underset{i_n=1}{\overset{i_n=0}{\gtrless}}
 \! | \mathbf{r}_{l,n}[3] |^2    +   
  | \mathbf{r}_{l,n}[4]  |^2  
$
in \cite{KiBlSa:14}. 
Following the same lines in \cite[pp.~217--218]{PrSal:07}, the corresponding conditional error probability of rule in Eq.~\eqref{eq:non_coh_detec_rule} is 
$ \frac{1}{2}\,\mathsf{e}^{  - \frac{\left(  a^{[\rm m]}_{l,n}  \right)^2   \, \mathtt{M}_n  \, \mathtt{SNR}^{[\rm m]}_{n}}{2\,(\mathtt{M}_n + 1) }  } $ 
and
$ \frac{1}{2}\,\mathsf{e}^{  - \frac{\left(  a^{[\rm b]}_{l,n}  \right)^2    \mathtt{SNR}^{[\rm b]}_{n}}{2 }  } $
for monostatic and multistatic systems, respectively. 
Note that the above expressions coincide  with the upper bounds of
conditional error  probabilities   given in Eq.~\eqref{eq:coherent_basic_monostatic}
and~\eqref{eq:cond_prob_error_bistatic}.
In other words,  the expressions 
in~\eqref{eq_prob_error_monostatic_upper_bound} and~\eqref{eq:proof_probability_of_error_bistatic}  
reflect  the
exact  BER performance  under  noncoherent envelope detection  in monostatic and 
multistatic architecture, respectively.\footnote{Results are connected with the signal
model of Section~II, which assumes perfect CFO estimation and correction (in the multistatic
case) and perfect DC blocking at the baseband signal for both monostatic and multistatic cases.}

\section{Information Outage Analysis}
\label{sec:outage}

Section~\ref{sec:sys_model} assumed that all switching (subcarrier) frequencies used by the tags  adhere to the orthogonality criterion  given in Eq.~\eqref{eq:criterion_non_coh}. 
In practice, that may not be always feasible due to frequency generation constraints (e.g., clock
drifts, lack of phase-locked loops). As a result, tags may be allocated to distinct subcarrier frequencies, which 
may cause adjacent channel interference. Due to the roundtrip nature of signal propagation in scatter radio (i.e.,
from illuminator to tag and back to reader), an adjacent in frequency tag $j$, i.e., tag with pair of
subcarrier frequencies $\left(F_{j,0}, F_{j,1}\right)$ relatively \emph{close} to $\left(F_{n,0}, F_{n,1}\right)$, 
may be received with significantly higher power than the tag of interest $n$; thus, any deviation from the 
orthogonality criterion may cause interference. Therefore, the relative spatial location of one tag versus the other, 
i.e., network topology,  as well as the subcarrier frequency allocation (denoted as $\mathcal{C}$) of $N$ available pairs of
subcarrier frequencies to $N$ tags does affect overall performance in practice. Notice that there are $N!$ possible subcarrier 
frequency assignments to $N$ tags.  In this section, set   $\mathcal{A}(n)\triangleq \mathcal{N} \backslash n$ denotes the tags that interfere the reception of
tag $n$, ${g}^{[\rm m]}_{l,n}=\left({a}^{[\rm m]}_{l,n}\right)^2 $ and ${g}^{[\rm b]}_{l,n}=\left({a}^{[\rm b]}_{l,n}\right)^2$. 
For simplification in the followed analysis, it is also assumed that $g_{l,n}$ and $g_{l,j}$ are statistically independent among different tags $(n \neq j)$, for 
{any} $l \in \mathcal{L}$.

\subsection{Monostatic}
For a given subcarrier frequency assignment $\mathcal{C}$,
incorporating imperfections as mentioned above, the \emph{instantaneous} signal-to-interference-plus-noise ratio (SINR) of tag $n$ 
  at the $l$-th time slot is given by:  
 \begin{align}
\text{SINR}^{\rm [m]}_{l,n}(\mathcal{C})  &\triangleq \frac{ 
 {g}^{[\rm m]}_{l,n}\,   \frac{\mathtt{M}_n}{\mathtt{M}_n + 1} \,  \mathtt{E}^{[\rm m]}_{ n} 
 }{\sum_{j \in \mathcal{A}(n)} 
  \rho_{nj}(\mathcal{C}) \, {g}^{[\rm m]}_{l,j}\,  \frac{\mathtt{M}_j}{\mathtt{M}_j + 1} \, \mathtt{E}^{[\rm m]}_{ j} 
+ {N}_0} ,
\label{eq:SINR_monostatic}
\end{align}
where parameter $\rho_{nj}(\mathcal{C})$ is inversely proportional to the  assigned subcarrier frequencies 
separation between tag $n$ and tag $j \in \mathcal{A}(n)$ \cite{VaBlLe:08}. It 
depends on the spectral efficiency of the specific binary modulation 
implemented at each tag and the filtering functions at the reader: 
\begin{equation}
\rho_{nj}(\mathcal{C}) = \max_{i_n, i_j \in \mathds{B} } \left \{ 
\left[\varepsilon_{n,j} \,  \left| F_{n,i_n}^{\mathcal{C}} -
 F_{j,i_j}^{\mathcal{C}}   \right| \right]^{-2} \right\},  ~ j \in \mathcal{A}(n),
\label{eq:max_rho}
\end{equation}
where $F_{n,i_n}^{\mathcal{C}}$ is the subcarrier frequency allocated under assignment $\mathcal{C}$ at tag $n$ for bit
$i_n \in \mathbb{B}$; parameter $\varepsilon_{n,j} $ is a constant that  depends on the modulation and pulse shaping used, as well as the mismatch between 
the clocks of tag $n$  and $j$. Subcarrier frequency difference raised at the second power, 
as opposed to the fourth power, is due to the power spectral density of FSK implemented at
 each tag, as opposed to (continuous phase) MSK \cite{VaBlLe:08}. 

The average received SINR  for any $(l,n) \in \mathcal{L}\times \mathcal{N}$  can be expressed as:
\begin{align}
 \mathtt{SINR}^{\rm [m]}_{n}(\mathcal{C}) &= 
 \frac{ \mathbb{E}\! \left[{g}^{\rm [m]}_{l,n}\right] \,  \frac{\mathtt{M}_n}{\mathtt{M}_n + 1} \,  \mathtt{E}^{[\rm m]}_{n}}{\sum_{j \in \mathcal{A}(n)} 
\rho_{nj}(\mathcal{C}) \, \mathbb{E}\! \left[{g}^{\rm [m]}_{l,j}\right]  \,  \frac{\mathtt{M}_j}{\mathtt{M}_j + 1} \, \mathtt{E}^{[\rm m]}_{j}  +   {N}_0} \nonumber  \\
&=
 \frac{    \mathtt{E}^{[\rm m]}_{ n}}{  \sum _{j \in \mathcal{A}(n)} 
\rho_{nj} (\mathcal{C})  \, \mathtt{E}^{[\rm m]}_{ j}  +   {N}_0}.
\label{eq:avg_SINR_monostatic}
\end{align}

\begin{proposition} \normalfont \label{prop:monostatic_outage_prob}
For fixed monostatic topology and given subcarrier frequency assignment  $\mathcal{C}$,
under Rayleigh fading, i.e.,  $\mathtt{M}_n= 1$,   outage probability for specific tag $n$ and given slot $l \in \mathcal{L}$, is upper bounded as follows:
\begin{equation}
 \mathbb{P}\!\left( \text{SINR}^{\rm [m]}_{l,n}(\mathcal{C}) \leq \theta   \right) \leq 
1 -  \mathsf{e}^{-  \sqrt{ \frac{  2  \theta}{ \mathtt{SINR}^{\rm [m]}_{n} (\mathcal{C}) } }  } .
\label{eq:bound_outage_monostatic_fixed_T_C}
\end{equation}
\begin{IEEEproof}
See  Appendix~\ref{Appendix:app3}.
\end{IEEEproof}
\end{proposition}
It is emphasized that the above outcome depends on both tag location and network topology (through $\mathtt{E}^{[\rm m]}_{ n}$),
as well as subcarrier frequency assignment $\mathcal{C}$ (through parameter $\rho(\mathcal{C})$ above).

\subsection{Multistatic}
For given subcarrier frequency channel assignment $\mathcal{C}$, the instantaneous SINR of tag $n$ for the $l$-th time slot is   
\begin{equation}
\text{SINR}^{\rm [b]}_{l,n} (\mathcal{C})  \triangleq \ \frac{{g}^{[\rm b]}_{l,n}\,  \mathtt{E}^{[\rm b]}_{l, n}}{\sum_{j \in \mathcal{A}(n)} 
\rho_{nj}(\mathcal{C}) \, g^{\rm [b]}_{l,j} \, \mathtt{E}^{\rm [b]}_{l, j}  +   {N}_0 }.
\label{eq:SINR_bistatic}
\end{equation}
The average SINR for multistatic case  for any $(l,n) \in \mathcal{L} \times \mathcal{N}$  is given by:
\begin{equation}
 \mathtt{SINR}^{\rm [b]}_{l,n} (\mathcal{C})  = 
 \frac{ \mathtt{E}^{[\rm b]}_{l, n}}{  \sum _{j \in \mathcal{A}(n)} 
\rho_{nj}  (\mathcal{C})   \, \mathtt{E}^{[\rm b]}_{l, j}  +   {N}_0}.
\label{eq:avg_SINR_bistatic}
\end{equation}

\begin{proposition} \normalfont \label{prop:bistatic_outage_prob}
For fixed multistatic topology and subcarrier frequency assignment  $\mathcal{C}$,
under Rayleigh fading, i.e., $\mathtt{M}_{ln}=\mathtt{M}_n=1$,  the outage probability for specific tag $n$ and a given slot $l \in \mathcal{L}$, is upper bounded as follows:
\begin{align}
 &\mathbb{P}\!\left( \text{SINR}^{\rm [b]}_{l,n}(\mathcal{C})  \leq \theta   \right)   \nonumber \\ 
 \leq &  1 - 2\, \sqrt{\frac{\theta}{\mathtt{SINR}^{\rm [b]}_{l,n}(\mathcal{C})}}
 \,\mathsf{K}_1\!\left(2 \sqrt{\frac{\theta}{\mathtt{SINR}^{\rm [b]}_{l,n}(\mathcal{C})}} \right),
\label{eq:bound_outage_bistatic_fixed_T_C}
\end{align}
where $\mathsf{K}_1(\cdot)$ is the first order modified Bessel function of the second kind \cite[Eq.~(10.27.3)]{Olver:10}.
\begin{IEEEproof}
The proof is provided in Appendix~\ref{Appendix:app4}.
\end{IEEEproof}
\end{proposition}

It is noted again that the above bound depends on tag location and network topology (through $\mathtt{E}^{[\rm b]}_{l, n}$), 
subcarrier frequency assignment $\mathcal{C}$ (through parameter $\rho(\mathcal{C})$ above) and slot $l$ (in contrast to the
monostatic case), since different CE per slot may be assumed.

\section{Diversity Reception \& Randomized Topologies}
\label{sec:div_rec_rand_tp}
\subsection{Diversity Reception}
For channel realizations changing  independently among different slots, i.e., if $g_{l,n}$ and $g_{l',n}$ are statistically independent for 
{any} $l\neq l' \in \mathcal{L}$, in addition to the assumptions followed so far, the following corollary regarding BER is relevant:
\begin{corollary} \normalfont
When each tag scatters the same information across the $L$ slots, BER for Rayleigh fading at the high SNR 
regime drops with $1/\mathrm{SNR}^d$, where  $d \geq L$ and $d=L/2$ for the multistatic and monostatic architecture, respectively. 
\begin{proof}
This stems from Propositions~\ref{prop:monostatic_div_order}, \ref{prop:bistatic_div_order} at Section~\ref{sec:BER}.
\end{proof}
\end{corollary}
Diversity receiver could exploit channel estimation and maximum ratio combining of the received SNRs across the $L$ slots
(coherent case) or selecting the maximum SNR across $L$ slots and performing detection there (noncoherent case). 

Diversity reception can be also applied in outage processing, briefly outlined below.
For a fixed subcarrier  frequency assignment $\mathcal{C}$,
the  {monostatic} information outage event  for the $n$-th tag over $L$ attempts (time
 slots)  is the probability that    $n$-th tag's SINR is below threshold $\theta$
over all $L$ slots, i.e.,
\begin{align}
 \mathbb{P} \left(   \bigcap_{l =1}^L\! \left\{\text{SINR}^{\rm [m]}_{l,n}(\mathcal{C})
  \leq \theta \right \} \! \right) 
 \! \overset{(a)}{=} \! &
  \left[ \mathbb{P}\!\left( \text{SINR}^{\rm [m]}_{l,n}(\mathcal{C}) \leq \theta   \right)
\right]^{L}   \label{eq:outage_prob_monostatic_2} \\
\overset{(b)}{\leq}   &
\left( 1 -  \mathsf{e}^{-  \sqrt{ \frac{  2  \theta}{ \mathtt{SINR}^{\rm [m]}_{n} (\mathcal{C}) } }  } \right)^L,
\label{eq:outage_prob_monostatic_3}
\end{align}
where $(a)$  is due to  the fact that $\{g^{\rm [m]}_{l,n}\}_{l \in \mathcal{L}}$ are IID across different $l$  and $(b)$ holds 
only under Rayleigh fading due to Eq.~\eqref{eq:bound_outage_monostatic_fixed_T_C}.

Similarly to the monostatic case, 
for fixed subcarrier frequency assignment $\mathcal{C}$  and given  threshold  $ {\theta}$,
 the  {multistatic} information outage probability of tag $n$ operating over $L$ time slots is
\begin{align}
&\mathbb{P}\left(   \bigcap_{l =1}^L\! \left\{\text{SINR}^{\rm [b]}_{l,n}(\mathcal{C})
  \leq \theta \right \} \! \right) \!
  \overset{(a)}{=} 
  \prod_{l =1}^L   \mathbb{P}\!\left( \text{SINR}^{\rm [b]}_{l,n}(\mathcal{C}) \leq \theta   \right)  \label{eq:outage_prob_bistatic_fixed_topology_assignment0} \\
  \stackrel{(b)}{\leq}&   \prod_{l=1}^L \left( 1 - 2\, \sqrt{\frac{\theta}{\mathtt{SINR}^{\rm [b]}_{l,n}(\mathcal{C})}}
 \,\mathsf{K}_1\!\left(2 \sqrt{\frac{\theta}{\mathtt{SINR}^{\rm [b]}_{l,n}(\mathcal{C})}} \right) \right),
\label{eq:outage_prob_bistatic_fixed_topology_assignment1}
\end{align}
where $(a)$ exploited the independence of 
$\left\{ \text{SINR}^{\rm [b]}_{l,n} (\mathcal{C}) \right \}_{l \in \mathcal{L}}$ across different $l$ and $(b)$ holds
only in a Rayleigh fading scenario due to Eq.~\eqref{eq:bound_outage_bistatic_fixed_T_C}.

\subsection{Average Probability over Random Square Grids}
\label{subsec:outage_prob_grid_networks}

In order to obtain
topology-independent outage probabilities,   expressions that
average  over possible topologies have to be obtained,  in order to average out the impact of the studied topology. 
 While there are infinite classes of topologies to choose,  our attention is  restricted to  the 
class of square $M\times M$ grid network topologies,
which have simple  two-dimensional (2D) representation and are easy to work with.
For simplicity, it is assumed that  grid resolution $\Delta$, determining the granularity of the square grid,
divides $M$, i.e., $M/\Delta = K \in \mathds{N}$ and the set of  2D square grid points is denoted as follows:  
\begin{equation}
 \mathcal{G}_{M,\Delta} = \left\{ \left[ k_1\Delta ~~  k_2\Delta \right]^{\top}: (k_1,k_2) \in \{0,1,\ldots,K\}^2 \right\} .
\label{eq:grid_net_example}
\end{equation}
Set $\mathcal{G}_{M,\Delta}$ has $(K+1)^2$ elements ($2$D grid points) and $M$, 
$\Delta$ are chosen such that $(K+1)^2 \geq N + L + 1$, i.e., possible tag locations are more than the total number of tags, emitters and reader. 
It is noticed that the larger the grid resolution is the more accurate becomes the evaluation of any studied outage metric.

For a monostatic architecture with fixed  SDR reader's position $\mathbf{u}_{\rm R} 
\in \mathcal{G}_{M,\Delta}$,
there are  $\mathtt{J}_{K,N}^{\rm [m]} \triangleq\binom{(K+1)^2 - 1}{N}$ ways to place 
$N$ tags in $\mathcal{G}_{M,\Delta} \backslash \{ \mathbf{u}_{\rm R} \}$.\footnote{If the ordering of how they are placed is not considered and the reader location is fixed.}
 This is the ensemble of admissible square grid monostatic topologies.  
The calculation of a topology-independent  average outage probabilities requires  averaging over all  
$\mathtt{J}_{K,N}^{\rm [m]}$ possible   topologies. Because  $\mathtt{J}_{K,N}^{\rm [m]}$ is practically enormous, especially for large $K$,
the averaging in this work is applied through Monte-Carlo, i.e.,  for a relatively large number 
of times, the following experiment is repeated:  uniformly and random select a topology
$ \mathcal{T}_{L}^{\rm [m]}$   from the ensemble of monostatic 
grid topologies 
(i.e., each of them has probability $1/\mathtt{J}_{K,N}^{\rm [m]}$) and estimate the outage probability for the sampled topology. 
Finally, averaging is applied over the sampled topologies. The same methodology is applied to a multistatic architecture with $L$ CEs and a single SDR reader, 
where $\mathtt{J}_{K,L, N}^{[\rm b]} \triangleq \binom{(K+1)^2-1-L}{N}=\binom{K^2+2K-L }{N}$  topologies exist, under the same assumption.
 Although the idea of averaging out over the ensemble of square grid topologies cannot offer   closed-form, expressions, 
it provides a simple, yet tractable method to evaluate any tag-topology-independent metric of interest through Monte Carlo.

\section{Energy Outage Probability for Passive Tags with RF Energy Harvesting}
\label{sec:harvesting}
In contrast to the semi-passive tags assumed so far,  passive tags do not have a dedicated (ambient or not) energy source. Instead,
passive tags harvest RF energy from the illuminating carrier(s). 
For completeness, this section extends the comparison of monostatic vs multistatic architectures with passive tags. 

For fixed position of CE, tag $n$ and SDR reader at the $l$-th time slot, the \emph{input} power at the 
RF harvesting circuit of tag $n$ is given by:  
\begin{align}
P_{{ \rm h},l,n}^{\rm [m]}  &= P_{{\rm R}} \,    \mathsf{L}_{ {\rm T}_n {\rm R} } \, ( a_{l, {\rm T}_n {\rm R} }  )^2,\\
P_{{ \rm h},l,n}^{ \rm [b]} &= P_{{\rm C}_l} \,    \mathsf{L}_{ {\rm C}_l {\rm T}_n } \, ( a_{l,{\rm C}_{l}{\rm T}_n})^2 ,
\end{align}
for monostatic and multistatic architecture, respectively.
Energy outage (EO) event at tag $n$ occurs when for all $L$ time slots the received input power at
tag $n$ RF   circuitry is below a threshold $\theta_{\rm h}$, which models:
a) the sensitivity of the RF harvesting circuit 
(i.e., the fact that RF harvesters offer zero power when input power is below a threshold) 
and b) the fact that all circuits require a specific minimum power to
operate.\footnote{State-of-the-art passive RFID tags exhibit $\theta_{\rm h}=-22$ dBm \cite{Du:16}.} Thus, the EO event is defined as follows:
\begin{align}
\! \mathbb{P}\!\left({\rm EO}_{L,n}^{\rm [m]}  \, \big| \theta_{\rm h} \! \right )  & \triangleq 
 \mathbb{P}  \left(   \bigcap_{l =1}^L\! \left\{ \! P_{{ \rm h},l,n}^{\rm [m]}  \!  \leq \theta_{\rm h} \! \right \} \! \right) \!,
 \label{eq:EO_monostatic}\\
\mathbb{P}\!\left({\rm EO}_{L,n}^{ \rm [b]}\,  \big| \theta_{\rm h} \! \right)   & \triangleq \mathbb{P}  \left(   \bigcap_{l =1}^L\! \left\{ \! P_{{ \rm h},l,n}^{\rm [b]} 
\!  \leq \theta_{\rm h} \! \right \} \! \right) \!, \label{eq:EO_bistatic} 
\end{align}
for monostatic and multistatic architecture, respectively. Network topology impact on the energy outage probabilities for multistatic and monostatic systems 
is present  due to the path-gains $\{ \mathsf{L}_{ {\rm C}_l {\rm T}_n } \}_{l \in \mathcal{L}}$ and 
  $\mathsf{L}_{ {\rm T}_n {\rm R} }$.

For the monostatic architecture, RVs $\{P_{{ \rm h},l,n}^{\rm [m]}\}_{l \in \mathcal{L}} $ are IID and each $P_{{ \rm h},l,n}^{\rm [m]} $
is Gamma RV with parameters  $\left(\mathtt{M}_n, \frac{P_{\rm R}    \mathsf{L}_{ {\rm T}_n {\rm R} }   }{  \mathtt{M}_n} \right)$ and  cumulative distribution 
function (CDF) given in \cite[Table I]{RoGo:08}; the monostatic energy outage event in~\eqref{eq:EO_monostatic} is given by: 
\begin{equation}
\mathbb{P}\!\left({\rm EO}_{L,n}^{\rm [m]} \, \big| \theta_{\rm h}\right )  =  \left(\frac{\sgamma \! \left( 
\mathtt{M}_n,\frac{\mathtt{M}_n    \theta_{\rm h}}{P_{\rm R}    \mathsf{L}_{ {\rm T}_n {\rm R} }       }\right) }{ \Gamma ( \mathtt{M}_n)} \right)^{L},
\end{equation}
where $\sgamma(a,x) = \int_{0}^x t^{a-1} \mathsf{e}^{-t} \mathsf{d}t$ is the lower incomplete gamma function.

For the multistatic architecture,  $P_{{ \rm h},l,n}^{\rm [b]}$ is
a Gamma RV with parameters  $\left(\mathtt{M}_{ln}, \frac{P_{{\rm C}_l}   \mathsf{L}_{ {\rm C}_l {\rm T}_n }   }{  \mathtt{M}_{ln}  } \right)$ 
and RVs $\{P_{{ \rm h},l,n}^{\rm [b]} \}_{l \in \mathcal{L}}$ are independent; as a result,  the  energy outage   event in~\eqref{eq:EO_bistatic}  is calculated as follows:
\begin{equation}
\mathbb{P}\!\left({\rm EO}_{L,n}^{ \rm [b]} \, \big| \theta_{\rm h} \right)  
=  
\prod_{l=1}^{L}\frac{\sgamma 
\! \left( 
\mathtt{M}_{ln},\frac{\mathtt{M}_{ln}    \theta_{\rm h}}{P_{{\rm C}_l}    \mathsf{L}_{ {\rm C}_l {\rm T}_n }    }\right) }{ \Gamma ( \mathtt{M}_{ln})}.
\label{energy_outage_bistatic}
\end{equation}

\begin{figure*}[!t]
\centering
\subfloat[]{\includegraphics[width = 0.85\columnwidth]{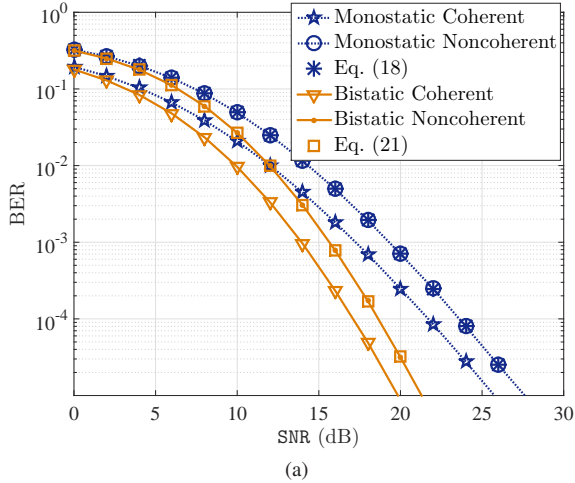}
\label{fig:BER_Monostatic_vs_Bistatic_equal_SNR_Kn10_Kln9_v2}} \quad
\subfloat[]{\includegraphics[width = 0.85\columnwidth]{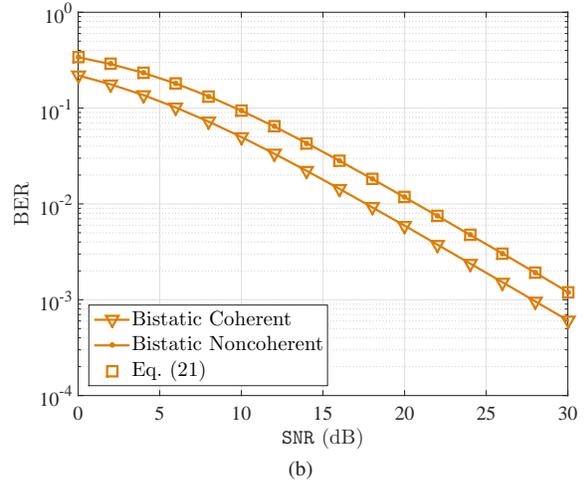}
\label{fig:BER_vs_SNR_Rayleigh_scenario_bistatic}}
\caption{Left (Right): Monostatic vs bistatic BER versus SNR for ML coherent and noncoherent detection and Nakagami fading
with parameters $\mathtt{M}_{n} =5.7619$, $\mathtt{M}_{ln} = 5.2632$  ($\mathtt{M}_{ln}=1, \mathtt{M}_{n} =5.7619$, assuming a NLoS CE-to-tag scenario).}
\label{fig:bistatic_vs_monostatic_BER}
\end{figure*}

The average across all tags energy outage event is by taking the average of Eqs.~\eqref{eq:EO_monostatic} and~\eqref{eq:EO_bistatic} 
across all tags, i.e.,
\begin{align}
\frac{1}{N} \sum_{n=1}^N  \mathbb{P}\!\left({\rm EO}_{L,n}^{\rm [m]} \, \big| \theta_{\rm h}\right ), ~\frac{1}{N} \sum_{n=1}^N \mathbb{P}\!\left({\rm EO}_{L,n}^{ \rm [b]} \, \big| \theta_{\rm h}\right)  
\label{eq:EO_avg} .
\end{align}
Another important metric that measures the worst-case energy outage  is
the maximum energy outage across all tags, i.e.,
\begin{align}
&  \max_{n \in \mathcal{N}} \left\{\mathbb{P}\!\left({\rm EO}_{L,n}^{\rm [m]} \, \big| \theta_{\rm h}\right )\right\},  ~\max_{n \in \mathcal{N}} \left\{ \mathbb{P}\!\left({\rm EO}_{L,n}^{ \rm [b]} \, \big| \theta_{\rm h}\right)   \right\}
\label{eq:EO_max} .
\end{align}

It is emphasized again that the above outage probabilities depend on a specific multistatic or 
monostatic WSN topology.   Energy outage expressions that are   topology-independent can be
offered using averaging over the ensemble of square grid topologies as in Section~\ref{subsec:outage_prob_grid_networks}.

\section{Simulation Results}
\label{sec:numerical_results}


First we study the BER performance of monostatic and bistatic  systems, for $L=1$ slot and $N=1$ tag. Rician fading parameters 
$\kappa_n =10$ and $\kappa_{ln} =9$ are considered, setting $\mathtt{M}_{n} =\frac{(\kappa_{n} + 1)^2}{2\kappa_{n}  + 1} = 5.7619$ and
$\mathtt{M}_{ln} =\frac{(\kappa_{ln} + 1)^2}{2\kappa_{ln}  + 1} = 5.2632$. Fig.~\ref{fig:BER_Monostatic_vs_Bistatic_equal_SNR_Kn10_Kln9_v2} illustrates the exact BER performance under
coherent and noncoherent reception for the two scatter radio architectures. Common $\mathtt{SNR}= \mathtt{SNR}^{[\rm b]}_{l,n} = 
\mathtt{SNR}^{[\rm m]}_{n}$ is assumed, resulting to monostatic reader transmission power 
$P_{\rm R}$ and bistatic carrier emitter power  $P_{{\rm C}_l}$ related according to  Eqs.~\eqref{eq:avg_energy_bistatic},~\eqref{eq:avg_energy_monostatic}
$\left(P_{\rm R} = \frac{\mathsf{L}_{{\rm C}_l,{\rm T}_n}}{
\mathsf{L}_{{\rm T}_n, {\rm R}}} \frac{\mathtt{M}_n   P_{{\rm C}_l}}{\mathtt{M}_n  +1}\right)$. 
Any monostatic or bistatic topology offering the specific SNR is applicable to this plot. Upper bounds derived
in Eq.~\eqref{eq_prob_error_monostatic_upper_bound}
and Eq.~\eqref{eq:proof_probability_of_error_bistatic} are also plotted. It can be seen that the bistatic
architecture outperforms the monostatic and high-SNR slope is clearly different among the two architectures, 
as expected.
It can be also seen that the derived upper bounds for coherent detection schemes  
coincide with the curves of noncoherent detection.


Fig.~\ref{fig:BER_vs_SNR_Rayleigh_scenario_bistatic} studies the BER performance
 as a function of SNR for a scenario  where the tag is far-away from CE (possibly a few kilometers) but the tag-to-reader distance is relatively
 small (a few feet). This is a Rayleigh-Nakagami (Rice) scenario, that could be the case when the tag is illuminated by broadcasting stations
 and the receiver is close to the tag. In such scenario, the monostatic architecture is not applicable. Due to the large 
 CE-to-tag distance, the link from CE-to-tag  is assumed NLoS,
 i.e., $\kappa_{ln} = 0$ and thus $\mathtt{M}_{ln} =\frac{(\kappa_{ln} + 1)^2}{2\kappa_{ln}  + 1} = 1$,
 while tag-to-reader is assumed strong LoS, thus $\kappa_{n} = 10$ with $\mathtt{M}_{n} =\frac{(\kappa_{n} + 1)^2}{2\kappa_{n}  + 1} = 5.7619$. 
 In such extreme case, where the monostatic architecture cannot be defined,
 the bistatic BER is less than $1\%$  for received SNR values
 less than $17$ dB ($20$ dB) under coherent (noncoherent) detection.
 The above demonstrate the potential benefits and flexibility of bistatic architecture
 in cases where CE is far away from tag.


\begin{figure}[!t]
\centering
\includegraphics[width = 0.85\columnwidth]{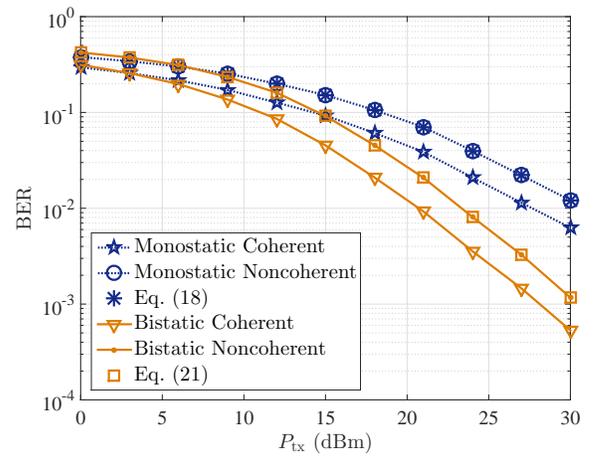}
\caption{Monostatic vs bistatic BER performance versus $P_{\rm tx}$ in a  $40 \times 40$ topology 
for ML coherent and noncoherent detection averaged over random tag locations.}
\label{fig:bistatic_vs_monostatic_BER_b}
\end{figure}

\begin{table}[!t]
\renewcommand{\arraystretch}{1}
\caption{Noise and Tag Parameters}
\label{ta:numerical_results}
\centering
\begin{tabular}{|c | c |  c|}      
\hline
${N_0} = -169$  dBm/Hz    & $F_{\rm c} = 868$ MHz  & $\lambda = \frac{3\cdot 10^{8}}{F_{\rm c}}$ m 
  \\
  \hline
  $|\Gamma_{n,0} - \Gamma_{n,1}|= 2$, $\forall n $ &  $\mathtt{s}_n = 0.1$, $\forall n $    &\\
  \hline
\end{tabular}
\end{table}

 \begin{figure*}[!t]
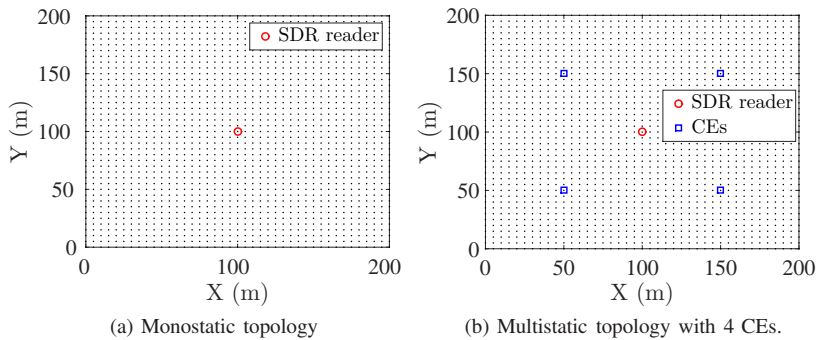

\centering
\subfloat[Monostatic topology]{\includegraphics[width = 0.6\columnwidth]{fig7.eps}
\label{fig:mon_topology}}
\subfloat[Multistatic topology with $4$ CEs.]{\includegraphics[width = 0.6\columnwidth]{fig8.eps}
\label{fig:multi_topology}}
\caption{Network setup:  Grid points (dots) are possible tag locations.}
\label{fig:bistatic_and_monostatic_topologies}
\end{figure*}

Fig.~\ref{fig:bistatic_vs_monostatic_BER_b} studies  the impact of transmit power on BER for random tag locations.
An SDR reader and a CE are placed in positions
$[0~0]^{\top}$ and $[40~40]^{\top}$, respectively,
while the position of tag follows uniform distribution over
a $40 \times 40$ m$^2$ topology. Common  transmit power is used for fair comparison for 
monostatic and bistatic architectures, i.e., $P_{\rm tx} = P_{\rm R} = P_{{\rm C}_l}$.
For each sampled tag location the  small scale fading parameters change as
$\kappa_n \sim \mathcal{U}[0,20]$ and $\kappa_{ln} \sim \mathcal{U}[0,20]$, setting $\mathtt{M}_{n} =\frac{(\kappa_{n} + 1)^2}{2\kappa_{n}  + 1}$ and
$\mathtt{M}_{ln} =\frac{(\kappa_{ln} + 1)^2}{2\kappa_{ln}  + 1}$.
In addition, the  path-loss exponents  (PLEs) from CE-to-tag 
and SDR reader-to-tag are $\mathcal{U}[2,2.5]$.
The noise- and tag-related parameters are shown in Table~\ref{ta:numerical_results}.
It can be seen that the average BER performance of
a randomly placed tag (evaluated over several possible tag locations) is smaller in the bistatic architecture. 
It is also observed that the BER decay is faster in the bistatic compared to the monostatic architecture, 
corroborating  the diversity gains offered by the bistatic system.

For grid WSN  topologies of size  $M\times M $ m$^2$ and grid resolution $\Delta$ m,  energy and information outage are examined, using $M=2.5$, $\Delta=0.125$ and $M=200$, $\Delta=5$, respectively.
 For information (energy) outage simulations, random tag topologies are generated from $\mathcal{G}_{200,5}$ ($\mathcal{G}_{2.5,0.125}$)
 consisting of $N$ tags, $L=4$ CEs 
placed at  $\{\mathbf{u}_{{\rm C}_l}\}_{l=1}^4 = \left\{ [\frac{M}{4}~~ \frac{M}{4}]^{\top}, 
[\frac{3M}{4}~~ \frac{M}{4}]^{\top}, [\frac{3M}{4}~~ \frac{3M}{4}]^{\top}, [\frac{M}{4}~~ \frac{3M}{4}]^{\top} \right\}$ 
 for multistatic system,
and an SDR reader placed at  $[\frac{M}{2}~~ \frac{M}{2}]$ (i.e., middle of the topology), in order to maximize the coverage. 
The grid topology utilized in information outage simulations is depicted in Figs.~\ref{fig:mon_topology} and~\ref{fig:multi_topology}
  for the monostatic and multistatic  architectures, respectively.  Similar with previous paragraph 
  noise/tag-related parameters are considered (Table~\ref{ta:numerical_results}).
 After sampling the grid
topology, random PLEs $\nu_{ln} \sim \mathcal{U}[2,2.5]$
and $\nu_{n} \sim \mathcal{U}[2,2.5]$ are generated for link from  $l$-th CE to $n$-th tag
and link from $n$-th tag to SDR reader, respectively.
In addition, to offer  robustness against channel fading, after sampling the topology
in the Nakagami fading scenarios, 
Nakagami  parameters are randomly generated as follows: $\mathtt{M}_{ln} \sim \mathcal{U}[1,5]$ and
$\mathtt{M}_{n} \sim \mathcal{U}[1,5]$, for link from  $l$-th CE to $n$-th tag
and link from $n$-th tag to SDR reader, respectively. 
For a fair comparison, equal transmission power
for monostatic and multistatic architecture is utilized, setting $P_{\rm tx} = P_{\rm R} =  P_{{\rm C}_l}, \forall l \in \mathcal{L}$.

\begin{figure}[!t]
\centering
\includegraphics[width = 0.85\columnwidth]{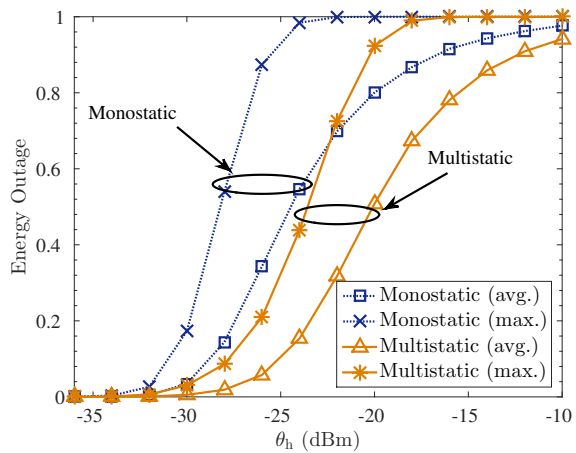}
\caption{Topology-independent average and maximum energy outage
performance versus threshold $\theta_{\rm h}$ for 
monostatic and multistatic network of Fig.~\ref{fig:bistatic_and_monostatic_topologies}
using  $\mathcal{G}_{2.5,0.125}$.}
\label{fig:EO_vs_theta_harvest_Ptx30_N100}
\end{figure}

Considering a passive-tag WSN scenario, Fig.~\ref{fig:EO_vs_theta_harvest_Ptx30_N100} examines
the topology-independent average and maximum energy outage probability 
(by averaging Eqs.~\eqref{eq:EO_avg} and~\eqref{eq:EO_max} over several
sampled grid topologies), as a function of harvesting threshold $\theta_{\rm h}$ under 
Nakagami fading, with $P_{\rm tx} = 35$ dBm and $N=8$ tags. It is noted that the probability for a tag to be placed near a CE is higher
in the multistatic architecture and thus, energy outage events are more frequent in
the monostatic architecture. Energy outage is a performance bound for WSNs consisting
of passive tags, since harvesting adequate energy is   necessary   before any other 
tag operation, including backscattering. 
It is remarked that for energy outage probability of $10 \%$, the multistatic architecture
outperforms the monostatic by $4.5$ dB for average and maximum outage performance.

\begin{figure}[!t]
\centering
\includegraphics[width = 0.85\columnwidth]{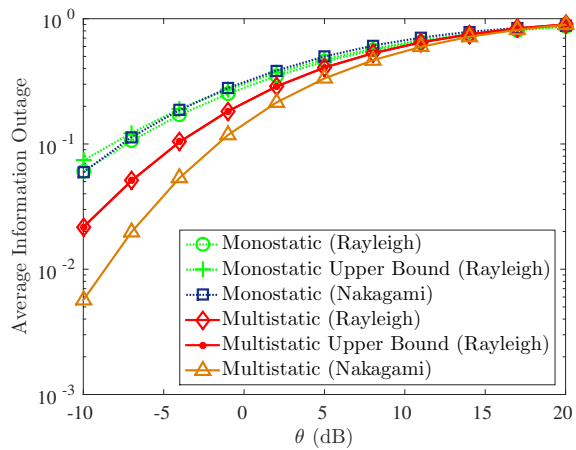}
\caption{Tag location-independent, average information outage probability versus threshold $\theta$ 
for monostatic and multistatic architecture of Fig.~\ref{fig:bistatic_and_monostatic_topologies}.}
\label{fig:bistatic_vs_monostatic_outage_theta}
\end{figure}
 
Finally, information outage is evaluated for a network of semi-passive tags.
  Since  average outage probability expressions in
Eqs.~\eqref{eq:outage_prob_monostatic_2} and~\eqref{eq:outage_prob_bistatic_fixed_topology_assignment0}  do not admit a closed form,  
an extra Monte Carlo step is required. Specifically, for a given sampled topology,
  Rayleigh $(\mathtt{M}_n=\mathtt{M}_{ln} =1)$ and Nakagami 
small-scale fading coefficients are generated, as well as a
random subcarrier frequency assignment to the tags. 
For the sampled topology, we average out the impact of fading and frequency assignment over several random realizations to evaluate 
 Eqs.~\eqref{eq:outage_prob_monostatic_2} and~\eqref{eq:outage_prob_bistatic_fixed_topology_assignment0}
through Monte Carlo.
The upper bounds for Rayleigh fading are evaluated directly through
Eqs.~\eqref{eq:outage_prob_monostatic_3}  and~\eqref{eq:outage_prob_bistatic_fixed_topology_assignment1}
after sampling the topology.
The above experiments are conducted for several sampled grid topologies 
to obtain  topology-free expressions.
 For an allocation $\mathcal{C}$, the $c$-th frequency pair is assigned
  to tag $n$ if $c = \mathsf{p}_{\mathcal{C}}(n) \in \mathcal{N}$, where $\mathsf{p}_{\mathcal{C}}$ 
  denotes the permutation associated with a specific assignment $\mathcal{C}$. The following parameters are utilized: $T = 1$ msec (i.e., $1$ Kbps bitrate), $\varepsilon_{n,j}=   2 \,\pi T $ for all $n,j \in \mathcal{N}$ with $n \neq j$, 
 and  the   subcarrier frequency pair for the $n$-th tag, $\left \{   F_{n,0}^{\mathcal{C}} ,  F_{n,1}^{\mathcal{C}} \right \}$,
is given by $F_{n,0}^{\mathcal{C}} = (0.1 + c\, F_{\rm sp}  )$ MHz and $F_{n,1}^{\mathcal{C}} = (0.1 + c\,F_{\rm sp} + F_{\rm sp}/5 ) $ MHz
with $F_{\rm sp} = 0.01$ MHz, $c=1,2,\ldots, N$. Thus, for a given channel assignment $\mathcal{C}$,
 the coefficients  in Eq.~\eqref{eq:max_rho} for any pair of tags $(n,j) \in \mathcal{N} \times \mathcal{N}$, with $n\neq j$,
can be expressed as $\rho_{nj}(\mathcal{C}) = \rho_{jn}(\mathcal{C})= \frac{25}{\left[2\,\pi \, T  \, (5|\mathsf{p}_{\mathcal{C}}(n)-\mathsf{p}_{\mathcal{C}}(j)|-1)\, F_{\rm sp}\right]^2}$.

Fig.~\ref{fig:bistatic_vs_monostatic_outage_theta} illustrates the topology-independent
average outage probability (as well as the corresponding upper bounds for Rayleigh fading)
for monostatic and multistatic architectures, as a function of threshold $\theta$, for $N=100$ tags
 and $ P_{\rm tx } = 28$ dBm. It is noted that the specific monostatic network setup (Fig.~\ref{fig:mon_topology})
is the most appropriate among  all possible choices of
$\mathcal{G}_{200,5}$ in terms of coverage, since the SDR reader is located in the middle of the grid. 
Fig.~\ref{fig:bistatic_vs_monostatic_outage_theta} shows that for  information outage $10\%$
multistatic system outperforms monostatic by $3$ dB in Rayleigh fading scenario, while for
Nakagami fading the gap approaches $8$ dB. The performance gap increases as threshold $\theta$
decreases. It can be also seen that the proposed bounds after averaging over all sampled tag locations are tight, especially
for the multistatic architecture. It is clear again that the multistatic architecture offers higher reliability, as well as better coverage for scatter radio WSNs.

\section{Experimental Results}
\label{sec:wsn}
%



 Motivated by the above finding, a digital \textit{multistatic} scatter radio sensor network is
 constructed \cite{Toun:16}, targeting environmental and agriculture applications for ultra low-cost, microclimate
 monitoring around each plant . The environmental quantities measured include air-humidity, soil moisture,
 temperature, which vary slowly with time and thus, low bitrate ($1$ Kbps) per sensor was adequate. Tags could 
 communicate with the reader simultaneously, using receiver-less subcarrier (switching) frequency division 
 multiple access (FDMA), as explained in Section II; the CEs transmitted with TDM access a CW at $13$~dBm power and $868$~MHz carrier frequency.

 \begin{figure}[!t]
\centering
\subfloat[Tag]{\includegraphics[width=2.0in]{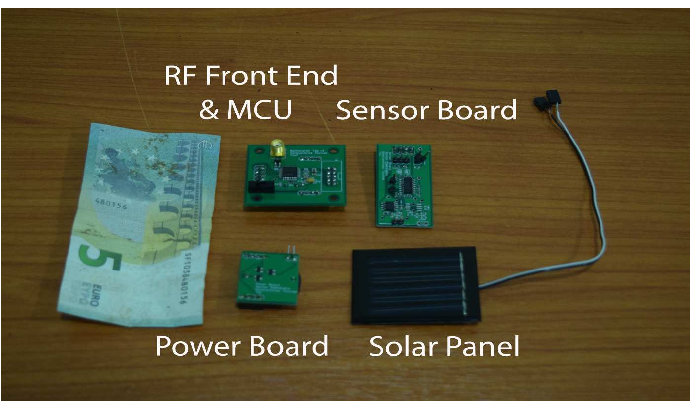}
\label{fig:scatter_radio_tag}
}
\subfloat[Reader]{
\includegraphics[scale=0.09]{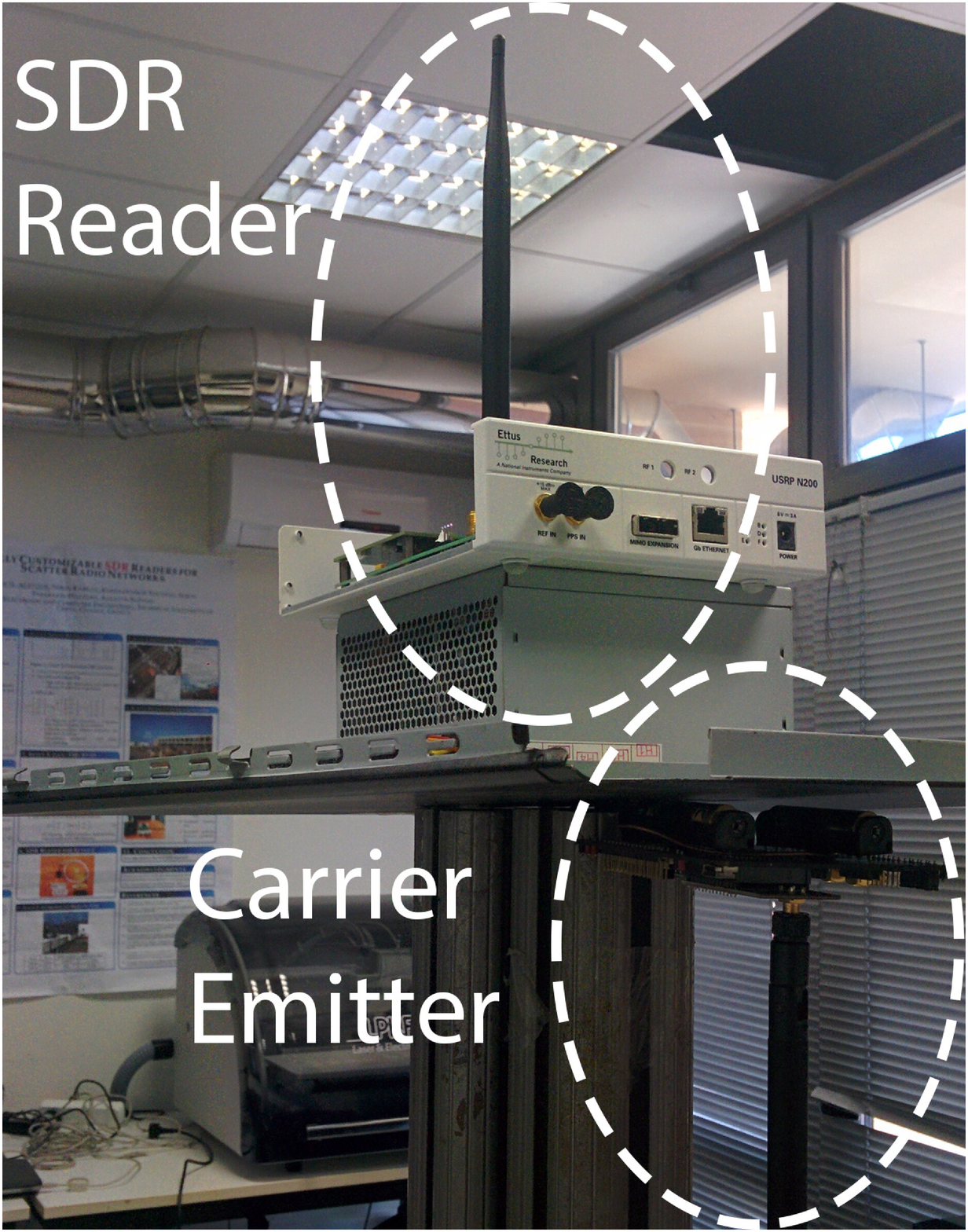}
\label{fig:monostatic_custom_reader}
}
\caption{A prototype scatter radio tag (left) and a 
custom reader (right); the receiver antenna is on the top side of a metal box, while the carrier emitter (CE) is on the bottom side.}
\end{figure}

Each prototype tag consisted of two distinct boards, the communication and the power board (Fig.~\ref{fig:scatter_radio_tag}). The communication board was responsible for the communication and sensing operations 
and included a $8$-bit mixed-signal micro-controller unit (MCU) with analog-to-digital converter, a RF transistor and input/output (I/O) pins for sensors' inputs.
Each tag utilized scatter radio binary FSK (BFSK) modulation; $30$ distinct orthogonal subcarrier frequency pairs
could be produced. In order to increase the total number of tags and sustain low-power operation, ``sleep" mode was implemented in all tags, with random
``wake up''. Thus, multiple tags can share the same pair of subcarrier frequencies, increasing the total number of tags in the network, based on the utilized 
``sleep-scatter'' duty cycle. Moreover, a $1/2$ rate Reed-Muller encoder was implemented at each MCU \cite{AlFaTouKaAgBl:14}. Finally, the tag operated with a
small solar panel during the daytime and a small coin cell 
battery during the night. The battery can be easily replaced by a super capacitor at the power board \cite{KaKiTouKoKoBl:14}, \cite{DaAsKaBle:14}, \cite{DaAsKaBl:16}.



Furthermore, in order to emulate a monostatic 
reader, the SDR reader and the CE were placed on the opposite sides of a metallic box. This structure provided good isolation  between the transmit and the 
receive antennas, without using a circulator. The monostatic setup is depicted in
Fig.~\ref{fig:monostatic_custom_reader}. 

\begin{figure*}[!t]
\centering
\subfloat[Multistatic: receiver at the corner, $3$ CEs.]{
\includegraphics[scale=.56]{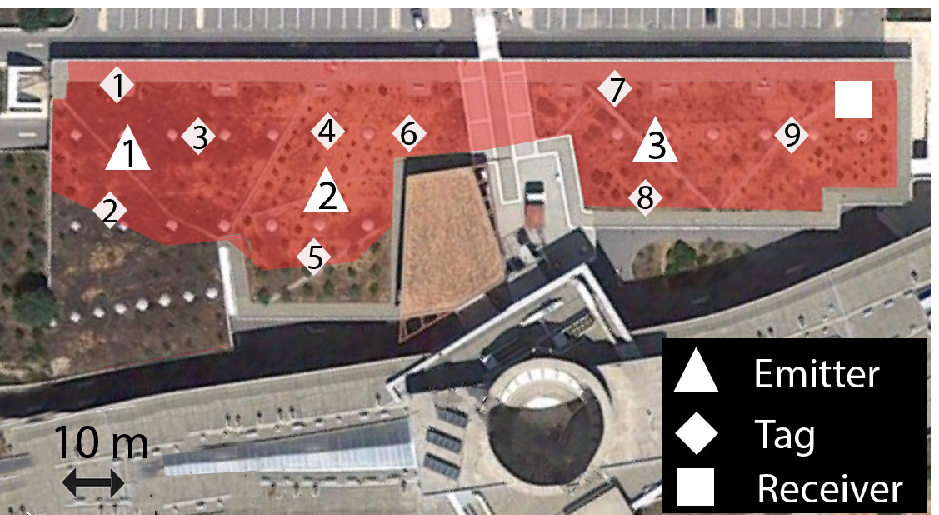}
\label{fig:bistatic_setup_1}
}
\subfloat[Multistatic: receiver at the center, $4$ CEs.]{
\includegraphics[scale=.56]{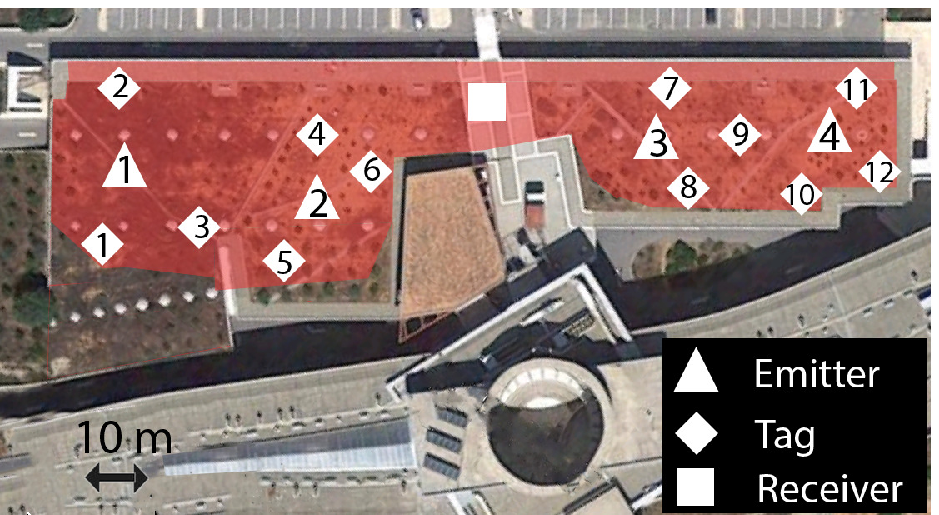}
\label{fig:bistatic_setup_2}
}
\subfloat[Monostatic: $8$ readers.]{
\includegraphics[scale=.56]{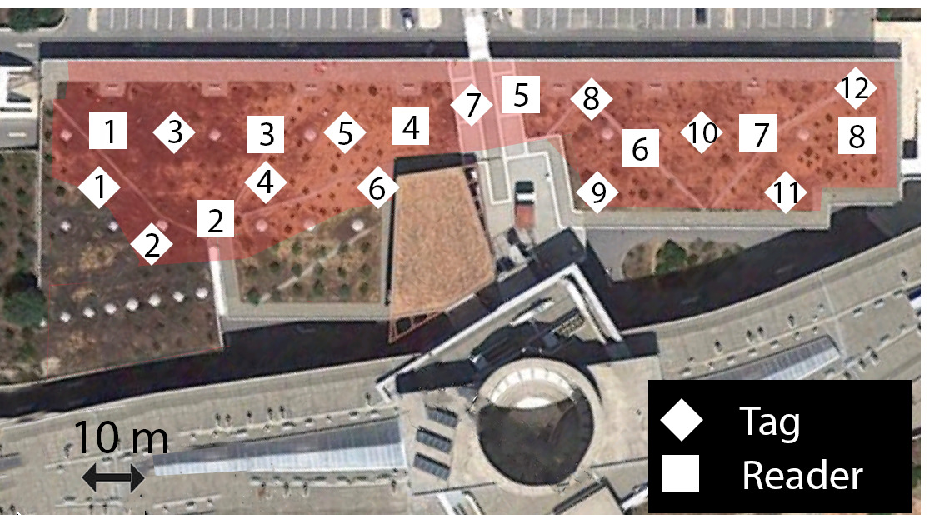}
\label{fig:monostatic_coverage}
}
\caption{Multistatic and monostatic setup for measurement campaigns.}
\end{figure*}

The outdoor measurement campaign consisted of two campaigns. In the first campaign, the maximum ranges of the tag-to-SDR  links were found. 
The metric used to find the maximum range was tag BER at the receiver, up to $5\%$. The multistatic architecture achieved tag-to-SDR reader 
ranges over $140$ m, with CE-to-tag range in the order of $10$ m.
The achieved ranges of the monostatic setup were one order of magnitude smaller than the multistatic, with maximum tag-to-SDR range in the order of $15$ m.

In the second campaign, coverage was examined in a field of $3500$ m$^2$ area. For the multistatic architecture two network topologies were deployed, one with three 
and one with four CEs. In the first topology (shown in Fig.~\ref{fig:bistatic_setup_1}), 
the SDR reader was placed at one field corner, maximizing the tag-to-SDR reader distance,
while the three CEs were placed around the field. 
In the second topology the reader was placed at the field center and the four CEs located as shown in Fig.~\ref{fig:bistatic_setup_2}. 
For the monostatic architecture, multiple readers/CEs were deployed. Since 
the monostatic architecture's ranges were shorter, a total of eight readers
were utilized, as shown in Fig.~\ref{fig:monostatic_coverage}. It is also noted that using only 
four monostatic SDR readers at the same locations, where CEs were placed in the multistatic case,
would offer a smaller monostatic coverage of $4 \cdot \pi\cdot 15^2=2827$ m$^2$.

The second campaign showed that both architectures can cover similar areas, provided that the monostatic architecture utilizes additional readers, incurring higher monetary cost.
In sharp contrast,   the multistatic architecture is more flexible with greater range, even with a single SDR receiver but multiple low-cost CEs. Equivalently, with only one reader, the monostatic architecture would offer much smaller field coverage, 
compared to the multistatic, as clearly shown in Fig.~\ref{fig:bistatic_setup_1}, Fig.~\ref{fig:bistatic_setup_2} and Fig.~\ref{fig:monostatic_coverage}.

The methodology of this work  can offer  tag-topology-independent performance metric expressions accounting
for the (random) spatial distribution of the tags over square grid topologies. Thus, the outcome of this work
could assist in assessing appropriate topologies for CEs and reader installation points,   with a simple, yet tractable way.
Specifically, in order to design backscatter sensor networks with increased range, extended information or energy coverage,
the following procedure can be applied:
a) create  a multistatic network consisting of a single reader placed in the middle of the topology and
$L$ carrier emitters (CEs) placed randomly over $L$ grid points; fix the location of the CEs and exclude $L+1$ grid points 
from the set of grid points that could be assigned to the tags; b) evaluate and store  
a tag-topology-free value for the metric of interest (BER or average information outage or average/maximum energy outage), 
averaged with respect to the positions of the $N$ tags over several realizations; c)
repeat the above steps $T_{\rm max}$ times, by changing each time the position of the $L$ CEs; 
d) among the  $T_{\rm max}$ possible CE topologies, choose the one  offering the smallest value for the metric of interest.
In that way, the system designer can offer convincing arguments for specific installation topology
of the reader and the CEs, exploiting the BER metric for flexible Nakagami or the outage probability metric for 
(worst-case) Rayleigh small-scale scaling. As a result, several multistatic topologies for the possible
location of the CEs or the reader can be assessed, depending on the application scenario and the wireless environment assumptions.
 Future research could extend the analysis of this work, in either monostatic or multistatic backscatter architectures, 
incorporating specific, application-dependent spatial distributions for the tags' locations.

\section{Conclusion}
\label{sec:conclusion}
It was clearly shown that the multistatic scatter radio architecture offers more reliable reception  as well as
better field coverage, while demonstrating smaller sensitivity to the topology of the tags, compared to 
the monostatic architecture. 
It is demonstrated under realistic Nakagami small-scale fading scenarios and 
path-loss models  that  not only the BER decay is doubled in multistatic architecture compared to monostatic one,
but also, energy  and information outage events are less frequent in multistatic systems
due to the flexible morphology of the multistatic WSN  architecture.
This work offers a concrete
proof  that large-scale and reliable scatter radio networks can be implemented, 
with digital, bistatic principles, and low-power scatter radio sensors.
%

\appendices

\section{Proof of Proposition~\ref{prop:monostatic_div_order}}

\label{Appendix:app1} %

For Rayleigh fading, i.e., $\mathtt{M}_n = 1$,  RV  $a^{[\rm m]}_{l,n}= \left(a_{ l,{\rm T}_n {\rm R} }\right)^2$
follows  exponential distribution. Using Eq.~\eqref{eq:coherent_basic_monostatic}, the unconditional probability of error over the $l$-th slot for the $n$-th tag is given by:
 \begin{align}
 {\rm Pr} \! \left(e^{\rm [m]}_{l,n} \right) =  &  \underset{a^{[\rm m]}_{l,n}}{\mathbb{E}} 
\left[ {\rm Pr} \left( e^{\rm [m]}_{l,n} \mid a^{[\rm m]}_{l,n}\right) \right]\! =\!\! 
\!\int_0^{\infty} \!\!\!\mathsf{Q}\!\left( x \sqrt{\frac{\mathtt{SNR}^{[\rm m]}_{n}}{2}}  \right) \mathsf{e}^{-x} \mathsf{d}x  \nonumber \\ 
= & \frac{1}{2} - \mathsf{e}^{ \frac{1}{  \mathtt{SNR}^{[\rm m]}_{n} }  } \mathsf{Q} \! \left(\sqrt{ \frac{2
}{\mathtt{SNR}^{[\rm m]}_{n}}} \right),
\label{eq_prob_error_monostatic}
\end{align}
where \cite[Eqs.~(7.14.2), (7.2.1), (7.2.2)]{Olver:10} are utilized. 
The diversity order for  the probability of error in~\eqref{eq_prob_error_monostatic}  is given by \cite{ZhThe:02}:
\begin{equation}
\lim_{\mathtt{SNR}^{[\rm m]}_{n} \to \infty}  \frac{ \log\!\left({\rm Pr} \! \left(e^{\rm [m]}_{l,n} \right)
\right)}{ \log\!\left(\mathtt{SNR}^{[\rm m]}_{n} \right)} 
=
 \lim_{x \to 0} \frac{\log\left( \frac{1}{2} - \mathsf{e}^x \,\mathsf{Q}\!\left( \sqrt{2x} \right) \right)}{\log\left( \frac{1}{x} \right)} .
\nonumber
\end{equation}
By applying the rule of L'Hospital:
\begin{equation}
 \lim_{x \to 0} \frac{-\mathsf{e}^x \mathsf{Q}\left(
\sqrt{2\, x} \right) +  \frac{1}{2 \sqrt{\pi \, x} }}{-\frac{1}{x} 
\left[\frac{1}{2} - \mathsf{e}^x \, \mathsf{Q}\left( \sqrt{2 \, x} \right)\right]} =
 \lim_{x \to 0} \frac{x \, \mathsf{Q}\left( \sqrt{2 \, x} \right) - 
\frac{\mathsf{e}^{-x} \sqrt{x}}{2 \sqrt{\pi}}}{\frac{\mathsf{e}^{-x}}{2} - \mathsf{Q}\left( \sqrt{2 \, x} \right)}.
\end{equation}
Applying again the rule of L'Hospital:
\begin{equation}
\lim_{x \to 0} \frac{\mathsf{Q}\left( \sqrt{2 \, x} \right) - 
\frac{\mathsf{e}^{-x}}{4{\sqrt{\pi x}}} }{-\frac{\mathsf{e}^{-x}}{2} + \frac{\mathsf{e}^{-x}}{2\sqrt{\pi x}}}
=\lim_{x \to 0} \frac{\mathsf{e}^x \sqrt{x} \, \mathsf{Q} \left( 
\sqrt{2 \ x} \right) - \frac{1}{4\sqrt{\pi}}}{-\frac{\sqrt{x}}{2} + \frac{1}{2 \, \sqrt{\pi}}} = -\frac{1}{2}. 
\label{eq:diversity_order_monostatic}
\end{equation}

%
\section{Proof of Proposition~\ref{prop:bistatic_div_order}}
\label{Appendix:app2} %

For Rayleigh fading, i.e., $\mathtt{M}_{ln} =\mathtt{M}_n = 1$,  with the aid of \cite[Eq.~(13.6.6)]{Olver:10}, function
$ \mathsf{U}\!\left(1, 1, \frac{2}{  \mathtt{SNR}^{[ \rm b]}_{l,n}} \right) =  \mathsf{e}^{ \frac{2}{  \mathtt{SNR}^{[ \rm b]}_{l,n}}}
\Gamma\!\left(0,  \frac{2}{  \mathtt{SNR}^{[b]}_{l,n}}\right)$, where 
$\Gamma(a,x)  \triangleq  \int_{x}^{\infty} t^{a-1} \, \mathsf{e}^{-t} \mathsf{d}t$ is the upper incomplete gamma function.
The diversity order for the upper bound  becomes   
\begin{align}
 &\lim_{x \to \infty} \frac{\log\left( \frac{1}{x} \, \mathsf{e}^{ \frac{2}{x} } \, \Gamma \! \left( 0, \frac{2}{x} \right)\right)}{\log\left(x\right)}  \nonumber \\
= & \lim_{x \to \infty} \left[ \frac{\log \left( \frac{1}{x} \right)}{\log (x)} + \frac{\log \left( \mathsf{e}^{ \frac{2}{x}} \right)}{\log (x)} + \frac{ \log \left( \Gamma \! \left( 0, \frac{2}{x}\right) \right)}{\log (x)}\right].
\label{eq:diversity_order_bistatic}
\end{align}
By applying the L'Hospital's rule and \cite[Eq.~(8.8.13)]{Olver:10}:
\begin{align}
 &\lim_{x \to \infty} \left[ \frac{\log \left( \frac{1}{x} \right)}{\log (x)} + \frac{2}{x\log (x)} + \frac{\mathsf{e}^{-\frac{2}{x}} }{ \Gamma \left( 0, \frac{2}{x} \right)} \right] 
 \nonumber \\
= &  -1 + \lim_{x \to \infty} \frac{\mathsf{e}^{-\frac{2}{x}} }{\Gamma\! \left( 0, \frac{2}{x} \right)}  =-1,
\end{align}
where $\lim_{x \to \infty} \frac{\log \left( \frac{1}{x} \right)}{\log (x)} = -1$, $\lim_{x \to \infty} \frac{2}{x\log (x)} = 0$ and
 $\lim_{x \to \infty} \frac{\mathsf{e}^{-\frac{2}{x}} }{\Gamma \left( 0, \frac{2}{x} \right)} = \lim_{x \to 0} \frac{1}
{\Gamma\! \left( 0, 2x \right)} = 0$, due to \cite[Eqs.~(8.4.4), (6.6.2)]{Olver:10}.

%
%
%

\section{Proof of Proposition~\ref{prop:monostatic_outage_prob}}
\label{Appendix:app3} %

The PDF of RV $ {g}^{[\rm m]}_{l,n}  \triangleq  \left(
 {a}^{[\rm m]}_{l,n} \right)^2 = \left( a_{ l,{\rm T}_n {\rm R} }\right)^4$ can be determined through  
RV $ {a}^{[\rm m]}_{l,n} = \left( a_{ l,{\rm T}_n {\rm R} }\right)^2$, which follows Gamma distribution with PDF given in Eq.~\eqref{eq:gamma_distr}.
Using the formula in \cite[p.~199]{pappoulis:02}, the required PDF for $x \geq 0$ is
\begin{equation}
\mathsf{f}_ {{g}^{[\rm m]}_{l,n}} (x) = \frac{1}{2 \sqrt{x}}  \mathsf{f}_ {{a}^{[\rm m]}_{l,n} }(\sqrt{x})  
=   (\mathtt{M}_{n})^{\mathtt{M}_{n}} \,
 \frac{x^{ \frac{\mathtt{M}_{n}}{2}- 1}}{2 \, \Gamma(\mathtt{M}_{n})} \, \mathsf{e}^{-\mathtt{M}_{n} \sqrt{x} }.  
\label{eq:monostatic_power_distr}
\end{equation}

For Rayleigh fading, i.e., $\mathtt{M}_n = 1$ in Eq.~\eqref{eq:monostatic_power_distr}, the PDF of RV
${g}^{[\rm m]}_{l,n}$ is simplified to $\mathsf{f}_{g^{[\rm m]}_{l,n}} (x  ) = \frac{1}{2\sqrt{x}} 
\mathsf{e}^{-\sqrt{x}}$,  $x \geq 0.$ The corresponding cumulative distribution function (CDF) is given by
\begin{equation}
\mathsf{F}_{{g}^{[\rm m]}_{l,n}} \!\left( x \right) =  \int_0^{x}  \mathsf{f}_{{g}^{[\rm m]}_{l,n}}\!\left(y\right) \mathsf{d} y = 1 - \mathsf{e}^{- \sqrt{x}}, \quad x \geq 0.
\label{eq:cdf_monostatic}
\end{equation}
The monostatic outage probability for $\mathtt{M}_n = 1$ follows:
\begin{align}
& \mathbb{P}\!\left(  \text{SINR}^{\rm [m]}_{l,n}(\mathcal{C}) \leq \theta   \right)   \overset{(a)}{=}
\mathbb{P}\!  \left(g^{[\rm m]}_{l,n}\leq \frac{2 \, \theta  \, N_0}
{\mathtt{E}^{[\rm m]}_{n}} + \frac{\theta \, {\rm I}^{[\rm m]}_{l,n}(\mathcal{C})}{ \mathtt{E}^{[\rm m]}_{n}} 
 \right)  \nonumber  \\
\overset{(b1)}{=}  & \underset{ {\rm I}^{[\rm m]}_{l,n}(\mathcal{C}) }{\mathbb{E}} \! \left[   \mathsf{F}_{g^{[\rm m]}_{l,n} | {\rm I}^{[\rm m]}_{l,n}(\mathcal{C})  }\! \!
\left( \frac{2  \,\theta  \,  {N}_0}{\mathtt{E}^{[\rm m]}_{n}} + 
\frac{ \theta \,   {\rm I}^{[\rm m]}_{l,n}(\mathcal{C})  }{\mathtt{E}^{[\rm m]}_{n} } \right) \right]    \nonumber \\
\overset{(b2)}{=}  & \underset{ {\rm I}^{[\rm m]}_{l,n}(\mathcal{C}) }{\mathbb{E}} \! \left[   \mathsf{F}_{g^{[\rm m]}_{l,n}  }\! \!
\left( \frac{2  \,\theta  \,  {N}_0}{\mathtt{E}^{[\rm m]}_{n}} + 
\frac{ \theta \,   {\rm I}^{[\rm m]}_{l,n}(\mathcal{C})  }{\mathtt{E}^{[\rm m]}_{n} } \right) \right]    \\
 \overset{(c)}{\leq}  &
\mathsf{F}_{g^{[\rm m]}_{l,n}  }\! \!
\left( \frac{2  \,\theta  \,  {N}_0}{\mathtt{E}^{[\rm m]}_{n}} + 
\frac{ \theta \, \mathbb{E}\! \left[ {\rm I}^{[\rm m]}_{l,n}(\mathcal{C}) \right]}{\mathtt{E}^{[\rm m]}_{n} } \right)  \nonumber \\
\overset{(d)}{=} & ~
 \mathsf{F}_{g^{[\rm m]}_{l,n}}\! \! \left( \frac{ 2 \,  \theta}{ 
 \mathtt{SINR}^{\rm [m]}_{n}(\mathcal{C}) }  \right)  =  1 -  \mathsf{e}^{-  \sqrt{ \frac{  2  \theta}{ \mathtt{SINR}^{\rm [m]}_{n}(\mathcal{C}) } }  }. 
\label{eq:upper_bound_outage_monostatic_3}
\end{align}
In $(a)$, total interference at tag $n$ for the monostatic system
 at the $l$-th time slot was defined as ${\rm I}^{[\rm m]}_{l,n}(\mathcal{C}) \triangleq  \sum_{j 
 \in \mathcal{A}(n)}\rho_{nj}(\mathcal{C})  \,g^{\rm [m]}_{l,j}\, 
   \mathtt{E}^{\rm [m]}_{j}$; step $(b1)$ exploited the law of iterated expectation \cite{Lev:08}; 
   step $(b2)$ exploited the assumed statistical independence between $g^{\rm [m]}_{l,n}$ and $g^{\rm [m]}_{l,j}$ for any $j\neq n $;
   step $(c)$ utilized  Jensen's inequality, taking into account the concavity of CDF in Eq.~\eqref{eq:cdf_monostatic}; $(d)$ exploited 
   the linearity of expectation, the SINR definition in Eq.~\eqref{eq:avg_SINR_monostatic}, and $\mathtt{M}_n = 1$.

\section{Proof of Proposition~\ref{prop:bistatic_outage_prob}}
\label{Appendix:app4} %

Proof  of Eq.~\eqref{eq:bound_outage_bistatic_fixed_T_C} is along the same lines with proof in
Proposition~\ref{prop:monostatic_outage_prob}. The CDF of ${g}^{[\rm b]}_{l,n}$ for Rayleigh fading
is needed, as well as proof of its concavity. Using $\mathtt{M}_{ln}=\mathtt{M}_n= 1$ in
Eq.~\eqref{eq:bistatic_power_distr}, the PDF is simplified to $\mathsf{f}_{{g}^{[\rm b]}_{l,n}}(x) = 2 
\,\mathsf{K}_0 \! \left(2 \sqrt{x} \right)$,  $ x \geq 0$. The corresponding CDF can be calculated as follows:
\begin{align}
\mathsf{F}_{{g}^{[\rm b]}_{l,n}}(x) &=\! \int_0^{x} \mathsf{f}_{{g}^{[\rm b]}_{l,n}}(y)
\mathsf{d}y = \int_0^{x} 2\,\mathsf{K}_0\!\left( 2 \sqrt{y} \right) \mathsf{d}y \nonumber \\
&\overset{(a)}{=}\int_0^{2\sqrt{x}} z\,\mathsf{K}_0\!\left( z \right) \mathsf{d}z 
\overset{(b)}{=} 4 x \int_0^{1} y \,\mathsf{K}_0\!\left( 2 \sqrt{x} \, y \right) \mathsf{d}y \nonumber \\
&\overset{(c)}{=} 1 - 2\, \sqrt{x} \,\mathsf{K}_1 (2 \sqrt{x} ), \, x \geq 0,
\end{align}
where step $(a)$ used $2\sqrt{y} = z$, $(b)$ $\frac{z}{2\sqrt{x}} = y$, and $(c)$ used \cite[Eq.~(6.561.8)]{gradshteyn:07}.
Concavity  of $\mathsf{F}_{{g}^{[\rm b]}_{l,n}}$ is shown by differentiating twice the CDF $\mathsf{F}_{{g}^{[\rm b]}_{l,n}}(x)$  and utilizing   \cite[ Eq.~(10.29.3)]{Olver:10}:
\begin{align}
\frac{\mathsf{d}^2}{\mathsf{d}x^2}\mathsf{ F}_{{g}^{[\rm b]}_{l,n}}(x) =\frac{\mathsf{d}}{\mathsf{d}x} \mathsf{f}_{{g}^{[\rm b]}_{l,n}}(x) &=
 \frac{\mathsf{d}}{\mathsf{d}x} 2 \, \mathsf{K}_0 \! \left(2 \sqrt{x} \right)  \nonumber \\
 &=  -\frac{2 \, \mathsf{K}_1( 2 \sqrt{x} )}{\sqrt{x}} < 0,
\end{align}
since $\mathsf{K}_1(x) > 0, \forall x > 0$. Thus,  $\mathsf{F}_{{g}^{[\rm b]}_{l,n}}$ is concave and Eq.~\eqref{eq:bound_outage_bistatic_fixed_T_C} follows 
using the same reasoning as in Proposition~\ref{prop:monostatic_outage_prob}.


\end{document}